\documentclass[12pt]{article}
\usepackage{graphicx}

\newcommand\pubnumber{SLAC--PUB--14352}
\newcommand\pubdate{January, 2011}

\def\SLAC{SLAC, Stanford University\\
    2575 Sand Hill Rd.,  Menlo Park, CA 94025  USA}
\def\doeack{\footnote{Work supported by the US Department of Energy,
                     contract DE--AC02--76SF00515.}}

\def\Title#1{\begin{center} {\Large #1 } \end{center}}
\def\Author#1{\begin{center}{ \sc #1} \end{center}}
\def\Address#1{\begin{center}{ \it #1} \end{center}}

\newcommand\pubblock{\rightline{\begin{tabular}{l} \pubnumber\\
         \pubdate \end{tabular}}}
\newenvironment{Abstract}{\begin{quotation} \begin{center}
                       ABSTRACT
     \end{center}\bigskip  }{\end{quotation}}
\newenvironment{Presented}{\begin{quotation} \begin{center} 
             PRESENTED AT\end{center}\bigskip 
      \begin{center}}{\end{center} \end{quotation}}

\def\Acknowledgements{\bigskip  \bigskip \begin{center} \begin{large}
             \bf ACKNOWLEDGEMENTS \end{large}\end{center}}

\def\beq{\begin{equation}}
\def\eeq#1{\label{#1}\end{equation}}
\def\eeqn{\end{equation}}

\newenvironment{Eqnarray}%
   {\arraycolsep 0.14em\begin{eqnarray}}{\end{eqnarray}}
\def\beqa{\begin{Eqnarray}}
\def\eeqa#1{\label{#1}\end{Eqnarray}}
\def\eeqan{\end{Eqnarray}}
\def\CR{\nonumber \\ }

\def\leqn#1{(\ref{#1})}

\let\bar=\overbar

\def\lsim{\mathrel{\raise.3ex\hbox{$<$\kern-.75em\lower1ex\hbox{$\sim$}}}}
\def\gsim{\mathrel{\raise.3ex\hbox{$>$\kern-.75em\lower1ex\hbox{$\sim$}}}}

\def\Re{{\rm Re}}

\def\L{{\cal L}}
\def\M{{\cal M}}

\def\tr{{\mbox{\rm tr}}}
\def\half{\frac{1}{2}}

\def\del{\partial}
\def\Dslash{\not{\hbox{\kern-4pt $D$}}}
\def\dslash{\not{\hbox{\kern-2pt $\del$}}}
\def\pslash{\not{\hbox{\kern-2pt $p$}}}
\def\qslash{\not{\hbox{\kern-2pt $q$}}}
\def\kslash{\not{\hbox{\kern-2pt $k$}}}
\def\oneslash{\not{\hbox{\kern-2pt $1$}}}
\def\twoslash{\not{\hbox{\kern-2pt $2$}}}
\def\threeslash{\not{\hbox{\kern-2pt $3$}}}
\def\fourslash{\not{\hbox{\kern-2pt $4$}}}

\def\ee{e^+e^-}

\def\msb{{\bar{\scriptsize M \kern -1pt S}}}

\def\drb{{\bar{\scriptsize D \kern -1pt R}}}

\def\eps{\epsilon}

\def\spa#1#2{ \langle #1 #2 \rangle }
\def\spb#1#2{ [ #1 #2 ] }
\def\apb#1 {  \langle #1 ] }
\def\bpa#1{  [ #1 \rangle }

\makeatletter
\def\section{\@startsection{section}{0}{\z@}{5.5ex plus .5ex minus
 1.5ex}{2.3ex plus .2ex}{\large\bf}}
\def\subsection{\@startsection{subsection}{1}{\z@}{3.5ex plus .5ex minus
 1.5ex}{1.3ex plus .2ex}{\normalsize\bf}}
\def\subsubsection{\@startsection{subsubsection}{2}{\z@}{-3.5ex plus
-1ex minus  -.2ex}{2.3ex plus .2ex}{\normalsize\sl}}

\renewcommand{\@makecaption}[2]{%
   \vskip 10pt
   \setbox\@tempboxa\hbox{\small #1: #2}
   \ifdim \wd\@tempboxa >\hsize     
       \small #1: #2\par          
     \else                        
       \hbox to\hsize{\hfil\box\@tempboxa\hfil}
   \fi}

 \def\citenum#1{{\def\@cite##1##2{##1}\cite{#1}}}
 
\newcount\@tempcntc
\def\@citex[#1]#2{\if@filesw\immediate\write\@auxout{\string\citation{#2}}\fi
  \@tempcnta\z@\@tempcntb\m@ne\def\@citea{}\@cite{\@for\@citeb:=#2\do
    {\@ifundefined
       {b@\@citeb}{\@citeo\@tempcntb\m@ne\@citea\def\@citea{,}{\bf ?}\@warning
       {Citation `\@citeb' on page \thepage \space undefined}}%
    {\setbox\z@\hbox{\global\@tempcntc0\csname b@\@citeb\endcsname\relax}%
     \ifnum\@tempcntc=\z@ \@citeo\@tempcntb\m@ne
       \@citea\def\@citea{,}\hbox{\csname b@\@citeb\endcsname}%
     \else
      \advance\@tempcntb\@ne
      \ifnum\@tempcntb=\@tempcntc
      \else\advance\@tempcntb\m@ne\@citeo
      \@tempcnta\@tempcntc\@tempcntb\@tempcntc\fi\fi}}\@citeo}{#1}}
\def\@citeo{\ifnum\@tempcnta>\@tempcntb\else\@citea\def\@citea{,}%
  \ifnum\@tempcnta=\@tempcntb\the\@tempcnta\else
  {\advance\@tempcnta\@ne\ifnum\@tempcnta=\@tempcntb \else\def\@citea{--}\fi
    \advance\@tempcnta\m@ne\the\@tempcnta\@citea\the\@tempcntb}\fi\fi}
\makeatother

\def\bfM{{\bf M}}

\begin{document}
\begin{titlepage}
\pubblock

\vfill
\Title{Simplifying Multi-Jet QCD Computation}
\vfill
\Author{Michael E. Peskin\doeack}
\Address{\SLAC}
\vfill
\begin{Abstract}
These lectures give a pedagogical discussion of the computation of 
QCD tree amplitudes for collider physics.  The tools reviewed are
spinor products, color ordering, 
MHV amplitudes, and the Britto-Cachazo-Feng-Witten recursion
formula.
\end{Abstract}
\vfill
\begin{Presented}
Mini-courses of the XIII Mexican School of Particles and Fields\\
University of Sonora, Hermosillo, Mexico \\ 
October 2--5, 2008 \\
 and \\ 
LHC Physics Summer School\\
Tsinghua University, Beijing, China\\
August, 16--20, 2010 
\end{Presented}
\vfill
\end{titlepage}
\def\thefootnote{\fnsymbol{footnote}}
\setcounter{footnote}{0}
\quad
\newpage

\tableofcontents

\newpage

\section{Introduction}

As the LHC begins operation, the number one topic of the day is QCD.
To a first approximation, everything that happens at the LHC is 
determined by QCD.  Events from new physics at the LHC will appear,
if we are lucky, at fractions
of a part per billion of the proton-proton total cross section. 
They will appear at rates ten thousand times smaller than
Drell-Yan processes, and a hundred times smaller than top quark pair
production. This  requires that we understand those Standard Model
 processes at a level
that includes
their dressing with QCD radiation.

Since the 1980's, there has been considerable work on methods for
QCD computation.  The primary goal of this work has been to make seemingly
impossible calculations doable. In the early 1980's, it was beyond the 
state of the art 
to compute to lowest order
the rates at hadron colliders for $W^+$ production with 
 three jets.   Recently, this process has been computed at the 1-loop
level, a computation that, if done by textbook methods, would have 
required thousands of Feynman diagrams, each one an integral over 
millions of terms~\cite{BlackHat,EKZ}.

A byproduct of this improved understanding of QCD computation is that
calculations of reasonable difficulty by textbook methods become
trivial when approached with the new methods.  In the LHC era, 
every graduate student ought to be able to
calculate the QCD amplitudes for multijet processes.  In this lecture, 
I will give you some tools to do that.

I will first review a set of ideas developed in the 1980's.  
Sections 2, 3, and 4 will develop, in turn, the ideas of {\it spinor
products}, {\it color ordering}, and {\it MHV amplitudes}.   In Section 
5, I will illustrate how these methods make the somewhat strenuous 
calculation of the most basic quark and gluon cross sections
 a triviality.  In Section 6,
I will review a relatively new wrinkle in this technology, the 
{\it Britto-Cachazo-Feng-Witten recursion formula}.  In Section 7, I will 
apply that formula to compute the 
Drell-Yan cross sections with one and two jets.  I hope that this last
example will give a persuasive illustration of the power of these 
methods.

There are many references for those who would like to learn more.
The first three topics are reviewed very beautifully in a 1991 Physics
Reports article by Mangano and Parke~\cite{ManganoParke} and in 
the 1995 TASI lectures by Dixon~\cite{DixonTASI}.  The newer set of 
calculational tools are described in recent review articles by 
Bern, Dixon, and Kosower~\cite{BDKreview} and Berger and 
Forde~\cite{BergerForde}.  These last articles are
especially concerned with technology that extends the ideas presented
here to the computation of loop amplitudes.

Loop calculations, though, are beyond the scope of these lectures.  My only
goal here is to do easy calculations.  I hope that these lectures will
help you extend the domain
of QCD calculations to which you will apply this term.

In parallel with learning these methods, it will be useful for you  to explore 
some of the available computer programs that automatically generate matrix
elements for multi-particle processes and also carry out Monte Carlo
integration over 
phase space, a topic that is not discussed in these lectures.  A particularly
accessible and powerful code is  MADGRAPH/MADEVENT~\cite{MGME,MGMEsite}.  
An important advanced code is SHERPA~\cite{SHERPA,SHERPAsite} 
with the matrix element
generator COMIX~\cite{COMIX}.
It is always good,
though, to know what is inside the black box and to have tools for extending
what is available there.  For that, I hope these lectures will provide
useful background.

\section{Spinor products}

Perturbative QCD is primarily concerned with the interactions of gluons
and quarks at momentum scales for which the masses of these particles
can be ignored.  It is well-known, and discussed in the 
textbooks, that massless particles can be labeled by their helicity.
For massless states, the helicity is a well-defined, Lorentz-invariant 
quantity.
The basic goal of these lectures will be to compute tree amplitudes
for massless quark and gluon states of definite helicity.

In the 1980's, Berends and Wu spearheaded an effort to compute amplitudes
for massless particles by exploiting the property that their lightlike
momentum vectors can be decomposed into spinors~\cite{BWu}.  Using this idea, 
scattering amplitudes can be written in terms of Lorentz-invariant 
contractions of spinors, {\it spinor products}. 

\subsection{Massless fermions} 

Consider a massless fermion of mommentum $p$.  The spinors for this fermion
satisfy the Dirac equation
\beq
      \pslash\ U(p) = 0
\eeq{Dirac}
There are two solutions to this equation, the spinors for right- and 
left-handed fermions.  In a basis where the Dirac matrices take the
form
\beq
       \gamma^\mu = \pmatrix{ 0 & \sigma^\mu \cr \bar\sigma^\mu & 0 \cr}\ ,
 \qquad   \gamma^5 = \pmatrix{-1 & 0\cr 0 & 1\cr}
\eeq{gammabasis}
where $\sigma^\mu = (1,\vec\sigma)$, $\bar\sigma^\mu = (1,-\vec \sigma)$,
these spinors take the form
\beq
      U_R(p) = \pmatrix{0 \cr u_R(p)\cr }\  ,
     \qquad   U_L(p) = \pmatrix{ u_L(p) \cr 0 \cr }
      \ , 
\eeq{twocompU}
where the entries are 2-component spinors satisfying
\beq
        p\cdot \sigma\ u_R = 0 \ , \qquad p\cdot \bar \sigma\ u_L = 0 \ .
\eeq{twocompDirac}
Each equation has one unique solution.  The spinor $u_R(p)$ can be 
related to the spinor $u_L(p)$ by 
\beq
           u_R(p) = i \sigma^2 u^*_L(p) \ ,
\eeq{RLrel}
since this transformation turns a solution of one of the equations 
\leqn{twocompDirac} into a solution to the other.  This equation also
gives a phase convention for $u_R(p)$ that I will use throughout these
lectures.

To describe antiparticles, we also need solutions $V(p)$ that describe
their creation and destruction.  However, for massless particles, 
$V(p)$ satisfies the same equation as $U(p)$.   We can then use
the same solutions for these quantities, with $u_R$ and $u_L$ related as above.
 The spinor $V_R(p)$ 
is used for the creation of a left-handed antifermion; the spinor $V_L(p)$
is used for the creation of a right-handed antifermion.

In the discussion to follow, it will be simplest to treat all particles
as final states of the amplitudes we are considering.  The outgoing
left- and right-handed fermions will be represented by the spinors
$\bar U_L$ and $\bar U_R$ and outgoing left- and 
right-handed antifermions will be
represented by the spinors $U_R$ and $U_L$.  I will recover initial-state
particles  by crossing, that is, by starting from the amplitude with
the antiparticle in the final state and evaluating that amplitude for 
a 4-momentum with negative time component.

I will represent these spinors compactly as
\beq
   \bar U_L(p) = \langle p \ , \quad \bar U_R(p) = [p \ , \quad
    U_L(p) =  p] \ , \quad  U_R(p) = p\rangle \ , 
\eeq{sqangle}
The Lorentz-invariant spinor products can then be standardized as
\beq
\bar U_L(p) U_R(q) = \spa pq\ ,   \qquad    \bar U_R(p) U_L(q) =  \spb  p q \ .
\eeq{spproddef}
The quantities on the right-hands sides are called simply {\it angle
brackets} and {\it square brackets}.
The spinors are related to their lightlike 4-vectors by the identities
\beq
   p\rangle [ p = U_R(p) \bar U_R(p) = \pslash ({1-\gamma^5\over 2}) \ ,
\quad   p]\langle p = U_L(p) \bar U_L(p) = \pslash ({1+\gamma^5\over 2})
\eeq{splitp}
From these formulae, we can derive some basic properties of the brackets.
First,
\beq
           \spa pq =  [qp]^* \ .
\eeq{conjsp}
Next, multiplying two matrices from \leqn{splitp} and taking the trace,
\beq
  \spa p q \spb qp =  \tr [ \qslash \pslash ({1+\gamma^5\over 2})] = 2p\cdot q
\eeq{sptodot}
so that
\beq
      | \spa p q |^2 = | \spb qp |^2 = 2 p\cdot q  \ .
\eeq{sptodotforsure}
Finally, using \leqn{RLrel},
\beq
    \spa pq = u^\dagger_L(p) u_R(q) =  u^*_{La}(p) (i\sigma^2)_{ab}
        u^*_{Lb}(q) \ .
\eeq{antisymmspa}
Then, by the antisymmetry of $\sigma^2$, 
\beq
     \spa p q = - \spa q p \, \qquad  \spb p q = - \spb q p \ .
\eeq{antisymmforsure}
We now see that the brackets are square roots of the corresponding Lorentz
vector products, and that they are antisymmetric in their two arguments.

Some further identities are useful in discussing vector currents built
from spinors.  First,
\beq
\bar U_L(p) \gamma^\mu U_L(q) = u_L^\dagger(p)\bar\sigma^\mu
 u_L(q) \ , \quad  \bar U_R(p) \gamma^\mu U_R(q) = u_R^\dagger(p)\sigma^\mu
 u_R(q) \ .
\eeq{basicUtou}
The  relationship between $u_L(p)$ and $u_R(p)$ in \leqn{RLrel} implies that
we can rearrange
\beqa
  u^\dagger_L(p) \bar\sigma^\mu u_L(q)
 &=& u^\dagger_L(p) \bar\sigma^\mu (- (i\sigma^2)^2 )u_L(q)\CR
 &=& u^\dagger_L(p)(-i\sigma^2) \sigma^{\mu T} (i\sigma^2) u_L(q)\CR
 &=& u_R^T(p) \sigma^{\mu T} u^*_R(q) \CR
 &=& u_R^\dagger(q) \sigma^{\mu} u_R(p)\ ,
\eeqa{reversesp}
From this, it follows that
\beq 
       \langle p \gamma^\mu q] = [ q \gamma^\mu p \rangle \ .
\eeq{reverseforsure}

The {\it Fierz identity}, the identity of sigma matrices
\beq
     (\bar \sigma^\mu)_{ab} (\bar \sigma_\mu)_{cd} = 2 (i\sigma^2)_{ac}
       (i\sigma^2)_{bd} \ ,
\eeq{firstFierz}
allows the simplification of contractions of spinor expressions.
In a matter similar to \leqn{reversesp}, the relation
\leqn{firstFierz} can be used to show
\beq
 \langle p \gamma^\mu q] \langle k \gamma_\mu \ell ] = 2 \spa pk \spb \ell q\ ,
 \qquad \langle p \gamma^\mu q] [ k \gamma_\mu \ell \rangle 
= 2 \spa p\ell \spb k q\ .
\eeq{Fierz}

Finally, spinor products obey the {\it Schouten identity}
\beqa
   \spa ij \spa k\ell + \spa ik \spa \ell j + \spa i\ell \spa jk   &=& 0 \CR
   \spb ij \spb k\ell + \spb ik \spb \ell j + \spb i\ell \spb jk   &=& 0 \ .
\eeqa{Schouten}
To prove this identity, note that
the expressions on the left
are totally antisymmetric in $j,k,\ell$. But, antisymmetrizing
three 2-component objects gives zero.

The identities presented in this section will allow us to reduce complex
spinor expressions to functions of angle brackets and square brackets, which
are the square roots of the Lorentz products of lightlike vectors.  In the
following, I will denote these Lorentz products using the notation
\beq
     s_{ij} = 2 p_i\cdot p_j = (p_i+ p_j)^2 \ .
\eeq{sijdefin}
To evaluate simple expressions written in terms of spinor products, 
we can often think of the whole expression as the square root of an
expression in terms of the $s_{ij}$.  However, for more complex expressions,
it is usually easiest to directly evaluate the spinor products from their
component momenta.  I give formulae for evaluating spinor
products in Appendix A.

\subsection{$\ee\to \mu^+\mu^-$}

\begin{figure}
\begin{center}
\includegraphics[height=1.2in]{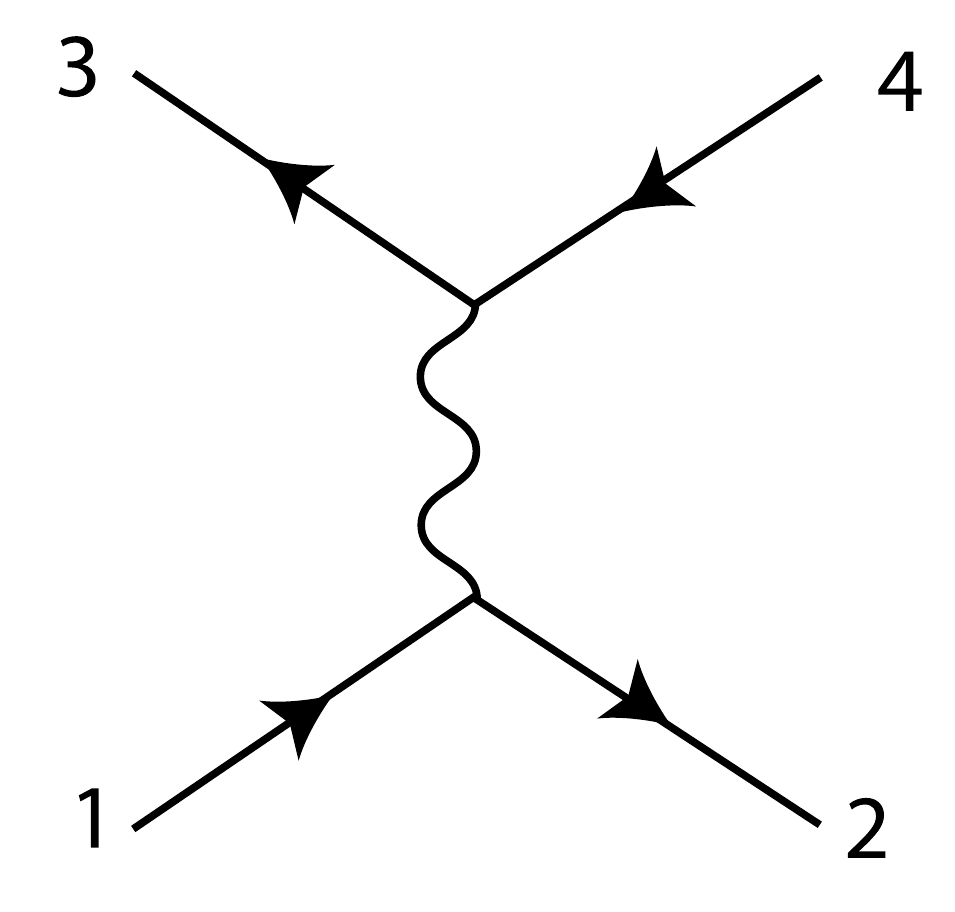}
\caption{Feynman diagram for $e_L^- e_R^+ \to \mu^-_L \mu^+_R$.}
\label{fig:eemumu}
\end{center}
\end{figure}
At this point, we are already able to perform some interesting computations.
Consider, for example, the tree-level
amplitude for $e_L^- e_R^+ \to \mu^-_L \mu^+_R$
in QED.  This is given by the diagram shown in Fig.~\ref{fig:eemumu}.
Label the momenta as in the diagram, considering all momenta as outgoing.
Then the amplitude is
\beqa
i\M &=& (-ie)^2 \ { -i\over q^2}\ \bar U_L(3) \gamma^\mu U_L(4) \, 
   \bar U_L(2) \gamma_\mu U_L(1) \CR    
& = &  { ie^2 \over q^2} \langle 3 \gamma^\mu 4]\langle 2 \gamma_\mu 1] \CR
     & = &  {2ie^2 \over q^2} \spa 3 2 \spb 14 \ ,
\eeqa{eemumu}
where I have used \leqn{Fierz} in the last step.
Now, $\spa 3 2$ and $\spb 14$ are both square roots of 
\beq
    s_{23} = (k_2 + k_3)^2 = (k_1 + k_4)^2 
\eeq{idtwothree}
which is just the Mandelstam invariant $u$. In the $\ee$ center of mass 
frame, $u = - 2E^2 (1 + \cos\theta)$, and $q^2 = s = 4E^2$.  Then
\beq
    |i\M|^2 =   e^4 (1 + \cos\theta)^2 \ .
\eeq{finaleemumu}
This is a familiar result, and we have obtained it with surprising ease.

Actually, we can simplify the result further.  In the denominator of 
\leqn{eemumu}, $q^2 = 2 k_1 \cdot k_2 = \spa 12 \spb 21$.  Multiply the
numerator and denominator of \leqn{eemumu} by $\spa 32$, to give
\beq
  i\M =  2ie^2 { \spa 32 \spb 14 \spa 32 \over \spa 12 \spb 21 \spa 32} \ .
\eeq{eemumutwo}
Then, in the denominator
\beq
 \spb 21 \spa 32 = 
  \spb 12 \spa 23 = [ 1 \twoslash\ 3 \rangle = [ 1 (-\oneslash -\threeslash
        - \fourslash) 3\rangle
\eeq{eemmsimple}
Using the Dirac equation, $\oneslash 1] = 0$, $\threeslash 3 \rangle = 0$.
This leaves $(- [14]\spa 43 )$.  The last square brackets cancel and we find
\beq
 i\M = 2ie^2  { {\spa 23}^2 \over \spa 12 \spa 34} \ .
\eeq{allangle}
The entire expression can be written in terms of angle brackets with 
no square brackets.  Similarly, if we had multiplied instead by 
\beq 
     { \spb 14 \over \spb 14 }
\eeq{allsquare}
a similar set of manipulations would have given
\beq
 i\M = 2ie^2 { {\spb 14}^2 \over \spb 12 \spb 34 } \ ,
\eeq{allsq}
with square brackets only.

\subsection{Massless photons}

This simplification of amplitudes with fermions extends to amplitudes with
massless vector bosons.  I will show that the polarization vectors for
final-state
massless vector bosons of definite helicity can be represented 
as~\cite{ChinesePol,GK,KS}
\beq
  \epsilon^{*\mu}_R(k) 
= {1\over \sqrt{2}} {\apb{r \gamma^\mu k} \over \spa r k }
  \ , \qquad  
  \epsilon^{*\mu}_L(k) 
= - {1\over \sqrt{2}}{\bpa { r \gamma^\mu k} \over \spb r k }
   \ .
\eeq{epsilons}
Here $k$ is the momentum of the vector boson and $r$ is some other fixed 
lightlike 4-vector, called the {\it reference vector}.  The only requirement
on $r$ is that it cannot be collinear with $k$.  

Here is the argument:  First, note that \leqn{epsilons}
satisfies the basic properties
\beq
      [ \eps^*_R(k) ]^* = \eps_L^*(k) \,  \qquad  
          k_\mu \eps^{*\mu}_{R,L}(k) = 0 \ .
\eeq{epsprops}
The second condition follows from $\kslash\, k] = 0$.  Also,
\beq
  \eps_R^*(k) \cdot [\eps_L^*(k)]^* = \eps_R^*(k) \cdot \eps_R^*(k) = 0 \ ,
\eeq{eRLorthog}
because, by \leqn{Fierz}, this expression is proportional to $\spa rr = 0$.
Finally,
\beq
  |\eps_R^*(k)|^2 = \half { \langle r {\gamma^\mu} k] 
       \langle k {\gamma_\mu} r ]\over
                 \spa rk \spb kr } = {2\over 2} { \spa r k \spb r k \over 
       \spa rk \spb kr } = -1 \ .
\eeq{sqtominusone}
Then 
\beq
     | \eps^*_R(k) |^2 = |\eps^*_L(k)|^2 = -1\ , \qquad 
  \eps^*_R(k) \cdot [ \eps^*_L(k) ]^* = 0\ .
\eeq{epsnorm}
as required for polarization vectors.

Next, evaluate the formulae for the $\eps^*_{R,L}$ for a particular choice
of the reference vector $r$.  For $  k = (k,0,0,k)  $, let $r = (r,0,0,-r)$.
The associated spinors are
\beq
   u_L(k) = \sqrt{2k}\pmatrix{0\cr 1\cr} \, 
\quad  u_R(k) = \sqrt{2k}\pmatrix{1\cr 0\cr} 
        \ , \quad 
         u_L(r) = \sqrt{2r}\pmatrix{-1\cr 0\cr}  \ .
\eeq{theusweneed}
Then 
\beqa
   \apb{ r {\gamma^\mu} k} &=&  
  u^\dagger_L(r) \bar \sigma^\mu u_L(k) = \sqrt{4kr}\ 
       (0, 1, -i, 0)^\mu  \CR
   \spa rk &=&  u^\dagger_L(r) u_L(k) =  - \sqrt{4kr}
\eeqa{useus}
For this choice of $r$, $\eps^*_R(k)$ is manifestly the right-handed
polarization vector,
\beq
    \eps^{*\mu}_R(k) =  - {1\over \sqrt{2}} (0,1, i, 0)^*
\eeq{idepsilon}

\begin{figure}
\begin{center}
\includegraphics[height=1.2in]{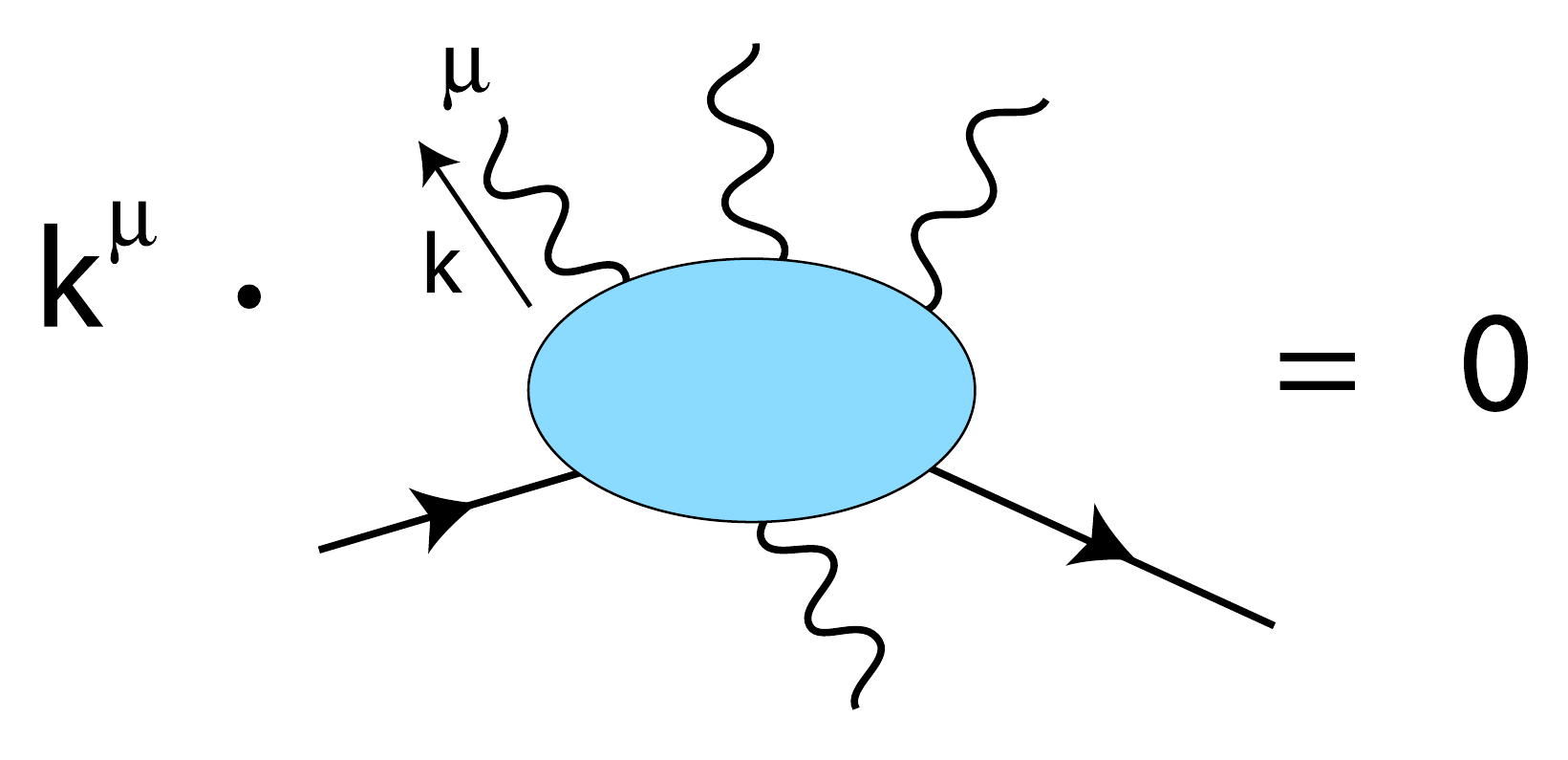}
\caption{Ward identity obeyed by a gauge-invariant sum of diagrams with
   all external particles on shell.}
\label{fig:Ward}
\end{center}
\end{figure}

Finally, analyze how \leqn{epsilons} changes when we change the
reference vector $r$ to a different lightlike vector $s$.
\beqa
   \eps^{*\mu}_R(k;r) - \eps^{*\mu}_R(k;s) &=&  {1\over \sqrt{2}} \left(
           {{ \apb{ r {\gamma^\mu} k } }\over \spa rk }  - 
          {{ \apb{  s {\gamma^\mu} k } }\over \spa sk } \right)\CR
    &=& {1\over \sqrt{2}} {1\over \spa rk \spa sk}(- \apb{r {\gamma^\mu} k}
      \spa ks +{ \apb{ s {\gamma^\mu} k } }\spa kr  ) \CR
   &=& {1\over \sqrt{2}} {1\over \spa rk \spa sk}(- 
           \langle r {\gamma^\mu} \kslash
      s \rangle + \langle s {\gamma^\mu} \kslash r \rangle ) \CR
   &=& {1\over \sqrt{2}} {1\over \spa rk \spa sk}\langle s 
    (\kslash \gamma^\mu +
{\gamma^\mu} \kslash ) r \rangle \CR
   &=&  {1\over \sqrt{2}} {1\over \spa rk \spa sk} \cdot 2 k^\mu \spa sr \ .
\eeqa{reducerseps}
The last line follows from the anticommutator of Dirac matrices.
The final result of this calculation is that
\beq
    \eps^{*\mu}_R(k;r) - \eps^{*\mu}_R(k;s)  =   f(r,s)  k^\mu \ .
\eeq{diffeps}
where $f(r,s)$ is a function of the two reference vectors.  This 
expression,  when dotted into an on-shell
 photon or gluon amplitude, will give zero by the Ward identity, as 
shown in Fig.~\ref{fig:Ward}.  Thus---as long as we are computing a 
gauge-invariant set of Feynman diagrams---we can use any convenient
reference vector $s$ and obtain the same answer as we would with the
particular reference vector $r$ used in \leqn{idepsilon}.

This completes the justification of the representation \leqn{epsilons}
of massless photon or gluon polarization vectors.  From here on, because
I will in any case consider all momenta as outgoing, I will drop the 
explicit $*$'s on the polarization vectors.

\subsection{$\ee\to \gamma\gamma$}

\begin{figure}
\begin{center}
\includegraphics[height=1.2in]{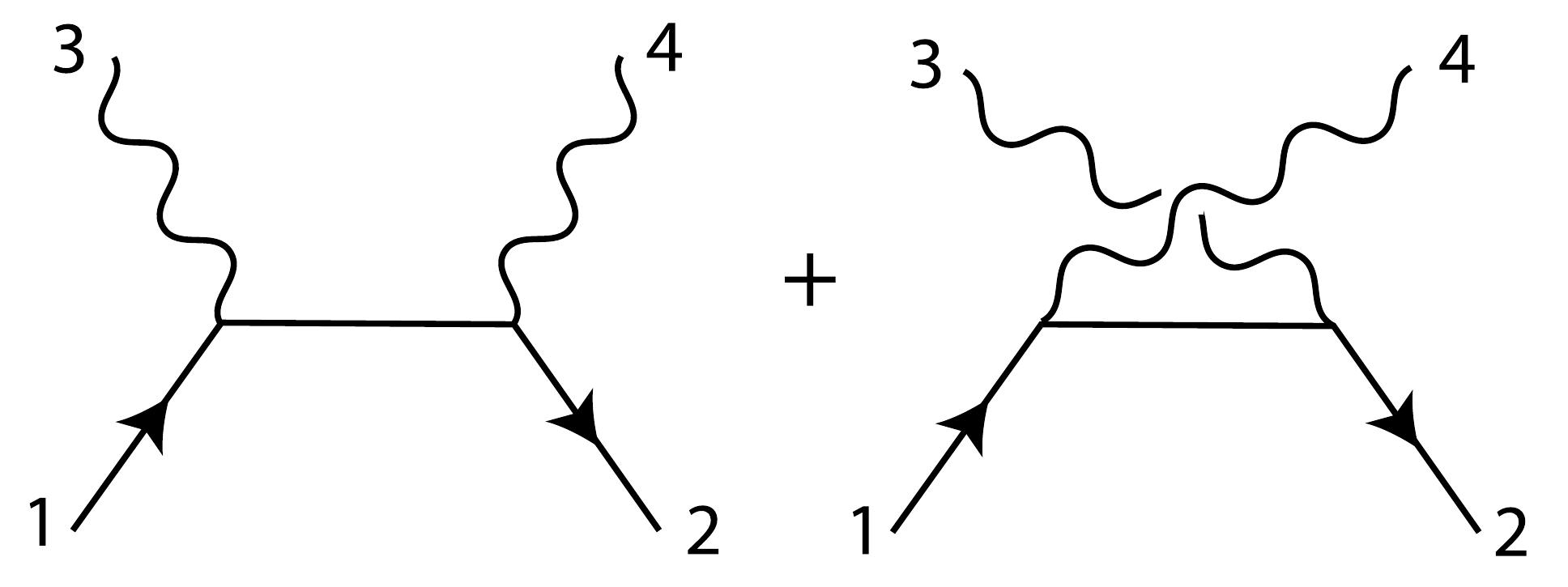}
\caption{Feynman diagram for $e_L^- e_R^+ \to \gamma\gamma$.}
\label{fig:eegammagamma}
\end{center}
\end{figure}

We can illustrate the application of these polarization vectors by 
computing the amplitudes for $e_L^- e_R^+ \to \gamma\gamma$.  Label 
the momenta as in Fig.~\ref{fig:eegammagamma}, taking all momenta as 
outgoing.  Then the value of the amplitude for this process is
\beq
    i\M = (-ie)^2 \langle 2 \left\{ \gamma\cdot \eps(4) { i (2+4)\over s_{24}}
    \gamma\cdot \eps(3) + \gamma\cdot\eps(3) { i (2+3) \over s_{23}} \gamma
       \cdot \eps(4) \right\} 1 ]\ . 
\eeq{eeggamplitude}
We will choose explicit polarization vectors $\eps(3)$,
$\eps(4)$ to evaluate this amplitude for each choice of photon helicities.
This formula also introduces
some additional streamlining of the notation:  I write $(2+4)$, instead
of $(\kslash_2 + \kslash_4)$ or even $(\twoslash + \fourslash)$, and I
use \leqn{sijdefin} to express the denominators.

There are four possible choices for the photon polarizations.  However
the cases $\gamma_R\gamma_L$ and $\gamma_L\gamma_R$ are related by interchange
of the momenta 3 and 4. The cases $\gamma_R\gamma_R$ and $\gamma_L\gamma_L$
are related by parity, which interchanges states with $R$ and $L$ polarization.
  Further, it is easy to see that the amplitudes
for $\gamma_R\gamma_R$ and $\gamma_L\gamma_L$ are actually zero. For the 
case of $\gamma_R\gamma_R$, choose $r = 2$ for both polarization vectors,
\beq
        \eps^\mu(3) = {1\over \sqrt{2}}{\apb{2 \gamma^\mu 3} \over \spa 23}
\ , \qquad  
 \eps^\mu(4) = {1\over \sqrt{2}}{\apb{2 {\gamma^\mu} 4} \over \spa 24}\ .
\eeq{epsRReegg}
When these choices are used in  \leqn{eeggamplitude}, we find, with the 
use of the Fierz identity \leqn{Fierz}
\beq
     \langle 2 \gamma \cdot \eps(4)\ \sim\  2 \spa 2 2 [4 \  =\  0 \ ,
\eeq{smashMeegg}
which vanishes
because $\spa 22 = 0$.  A similar cancellation occurs with 
$\eps(3)$.   So the entire matrix
element vanishes.  The  amplitude for the 
case $\gamma_L\gamma_L$ must then also vanish by parity; alternatively, 
we can find the same 
cancellation for that case by using $r = 1$ in both polarization vectors.

To compute the amplitude for the case $\gamma_R\gamma_L$, choose
\beq
        \eps^\mu(3) = {1\over \sqrt{2}}{\apb{2 \gamma^\mu 3} \over \spa 23}
\ , \qquad  
 \eps^\mu(4) = -{1\over \sqrt{2}}{\bpa{ 1 {\gamma^\mu} 4 }\over \spb 14}\ .
\eeq{epsRLeegg}
Then the second diagram in Fig.~\ref{fig:eegammagamma} vanishes by the 
logic of the previous paragraph.  Using the Fierz identity, the 
 first diagram gives
\beqa
    i\M &=& {-ie^2 \over s_{24} }{ 2\cdot 2 \over (-2) \spa 23 \spb 14}
 \spa 24 [ 1 (2+4) 2 \rangle \spb 31 \CR
 &=& {2 ie^2 \over s_{13} \spa 23 \spb 14}
 \spa 24  \spb 14 \spa 42 \spb 31 \CR
 &=& {2 ie^2 \over \spa 13 \spb 31 \spa 23 \spb 14}
  \spa 24  \spb 14 \spa 42 \spb 31 \CR
 &=&  2 ie^2 { (\spa 24)^2  \over
  \spa 23 \spa 31 }   
\eeqa{eeggRL}
In terms of the Mandelstam variables, $s_{23}= u$, $s_{13} = s_{24} = t$,
so
\beq
  |i\M|^2 =  4e^4 {t\over u}  =   4 e^4 { 1-\cos\theta\over 1 + \cos\theta}\ ,
\eeq{eeggRLresult}
which is the well-known correct answer.
In this case, the calculation gave directly the simplified form of the 
expression that involves only angle brackets and no square brackets.

The methods of this section can be generalized to massive external 
particles.  For the treatment of $W$ and $Z$ bosons, see Section 7.  
An efficient formalism for the treatment of massive fermions is presented
in~\cite{SW}.

\section{Color-ordered amplitudes}

We could now apply this spinor product technology directly to QCD.
However, it will be useful first to spend a bit of effort analyzing the
color structure of QCD amplitudes.  It is convenient to divide
QCD amplitudes into irreducible, gauge-invariant components of definite
color structure.  We will see that it is most straightforward to
  compute these objects separately 
and then recombine them
to obtain the full QCD results.

\subsection{$q\bar q \to gg $}

\begin{figure}
\begin{center}
\includegraphics[height=1.2in]{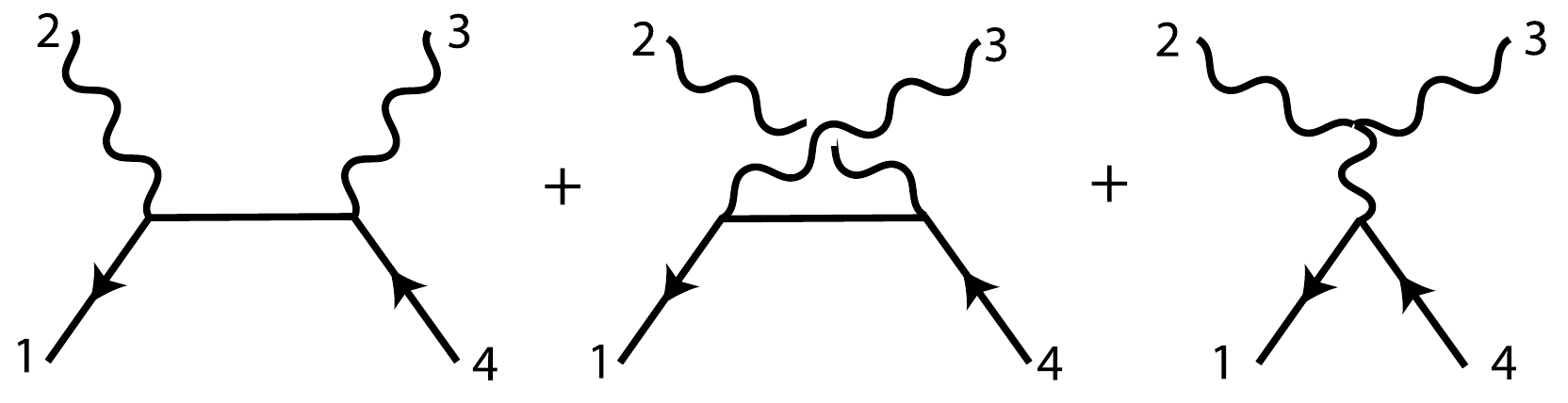}
\caption{Feynman diagrams for $q_L \bar q_R \to gg$.}
\label{fig:qqgg}
\end{center}
\end{figure}

Begin with the process $q_L \bar q_R \to gg$.  This is similar to the 
process $\ee\to \gamma\gamma$ analyzed in the previous section, except
that now there are three diagrams, as shown in Fig.~\ref{fig:qqgg}, including
one with a 3-gluon vertex.  With the numbering of external 
states as in the figure and all momenta outgoing, the value of the 
amplitude is
\beqa
    i\M &=& (ig)^2 \langle 1 \left\{ \gamma\cdot \eps(2) { i (1+2)\over s_{12}}
    t^a t^b
    \gamma\cdot \eps(3) + \gamma\cdot\eps(3) { i (1+3) \over s_{13}}t^b t^a
             \gamma
       \cdot \eps(2) \right\} 4 ]\CR
 & & + (ig) (-g f^{abc}t^c) {-i\over s_{14}}
      \langle 1 {\gamma^\lambda} 4] \cdot
    \biggl(\eps(2) \cdot \eps(3) (2-3)_\lambda \CR 
 & & \hskip 0.5in + \eps_\lambda(3) (2 \cdot 3 +2)\cdot
 \eps(2) +  \eps_\lambda(2) (-2 \cdot 2 -3)\cdot
 \eps(3) \biggr) \ .
\eeqa{qqggamplitude}
In this formula, $t^a$ and $t^b$ are the color $SU(3)$ representation matrices
coupling to the gluons 2 and 3, respectively.  The third diagram in 
Fig.~\ref{fig:qqgg} has a color structure that can be brought into the 
forms seen in the first two diagrams by writing
\beq
    -g f^{abc} t^c = (ig) \cdot if^{abc}t^c =( ig) (t^a t^b - t^b t^a) \ .
\eeq{changef}
We will find it convenient to rescale the color matrices: $T^a = \sqrt{2}t^a$,
so that the $T^a$ are normalized to 
\beq
     \tr [T^a T^b] = \delta^{ab} \ .
\eeq{Ttrace}
We can then write the amplitude in \leqn{qqggamplitude} in the form
\beq
  i\M = i \bfM(1234) \cdot T^a T^b +  i \bfM(1324) \cdot T^b T^a \ ,
\eeq{colorDecqqgg}
with 
\beqa
  i \bfM(1234) &=& ({ig\over \sqrt{2}})^2 \biggl[ \langle 1 \gamma\cdot \eps(2)
   { i (1+2)\over s_{12}} \gamma\cdot\eps(3) 4 ] \CR
& & \hskip 1.0cm + {-i\over s_{23}} \langle 1 \gamma^\lambda 4] 
   [\eps(2) \cdot \eps(3) (2-3)_\lambda \CR
 & & \hskip 2.5cm + \eps_\lambda(3) (2 \cdot 3 +2)\cdot
 \eps(2) +  \eps_\lambda(2) (-2 \cdot 2 -3)\cdot
 \eps(3) ] \ .
\eeqa{colororderedqqgg}
In the second term of \leqn{colorDecqqgg}, $\bfM(1324)$ is given by the 
same expression with $(2,\eps(2))$ exchanged with $(3,\eps(3))$.

The elements $i\bfM$ are called {\it color-ordered amplitudes}.  The 
complete helicity amplitudes such as \leqn{colorDecqqgg} are gauge-invariant
for any choice of helicities of the external particles and for any 
Yang-Mills gauge group.  The two color factors $T^aT^b$ and $T^bT^a$ are 
independent in any non-Abelian gauge group.  Thus, the color-ordered
amplitudes must be separately gauge-invariant.  This important observation
means that we can apply all of the simplifications discussed in the 
previous section to individual 
color-ordered amplitudes.  It will be much simpler to work with these
objects rather than with the full QCD amplitudes.

A consequence of this idea is that the individual color-ordered 
amplitudes should obey the Ward identity.  It is not difficult to 
verify this for \leqn{colororderedqqgg}.   Replace
$\eps(2)$  by the four-vector 2, and use 
the properties that the external momenta are lightlike, that $\langle 1$
and $4]$ satisfy the Dirac equation, and that  $3\cdot \eps(3) = 0$.
Then all terms in the resulting expression cancel.

\begin{figure}
\begin{center}
\includegraphics[height=1.0in]{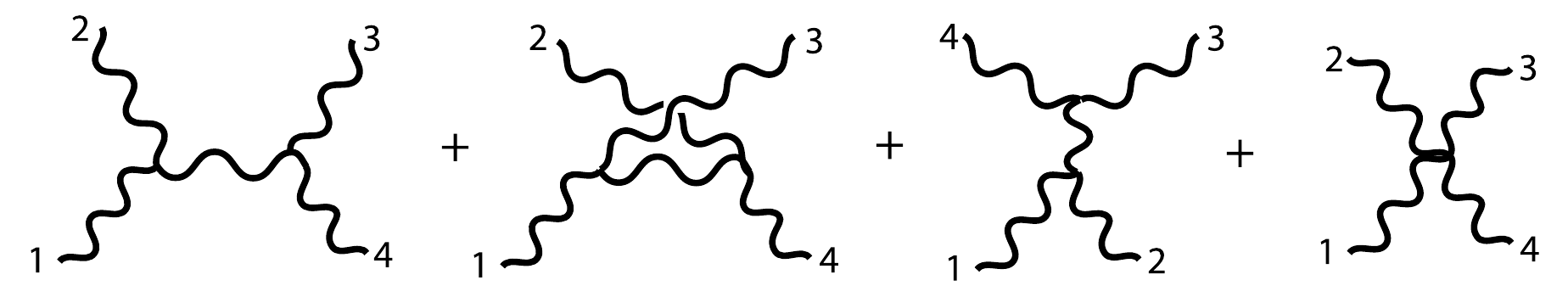}
\caption{Feynman diagrams for $gg\to gg$.}
\label{fig:gggg}
\end{center}
\end{figure}

Other QCD amplitudes can also be reduced to color-ordered structures.  Another
case that will be important for us is the 4-gluon amplitude shown in
Fig.~\ref{fig:gggg}.  This amplitude can be written in the form
\beqa
  i\M &=& i\bfM(1234)\cdot \tr[T^aT^bT^cT^d] 
+ i\bfM(1243)\cdot \tr[T^aT^bT^dT^c]\CR
 & &  +  i\bfM(1324)\cdot \tr[T^aT^cT^bT^d] 
 + i\bfM(1342)\cdot \tr[T^aT^cT^dT^b] \CR  & & 
+ i\bfM(1423)\cdot \tr[T^aT^dT^bT^c] +  i\bfM(1432)\cdot \tr[T^aT^dT^cT^b]\ .
\eeqa{ggggcolorDec}
I have used the cyclic invariance of the trace to move $T^a$ to the front
in each trace.  Then there are $3! = 6$ possible traces, all of which 
should be included.
For a sufficiently large $SU(N)$ gauge group, all of these traces are 
independent; thus, each coefficient is a gauge-invariant structure.
As in the case of $q\bar q gg$, these coefficients are given by a 
single function evaluated with different permutations of the external 
momenta.

\begin{figure}
\begin{center}
\includegraphics[height=1.0in]{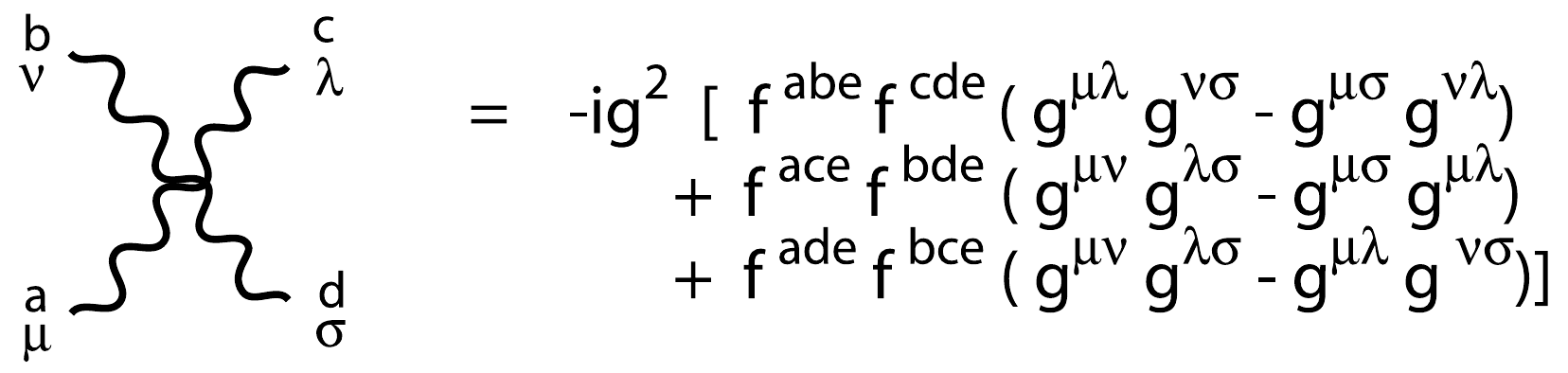}
\caption{The four gluon vertex of QCD.}
\label{fig:fourglue}
\end{center}
\end{figure}

To write the four diagrams in Fig.~\ref{fig:gggg} as a sum of color
structures, we need to convert color factors in 
the 3- and 4-gluon vertices into products
of $T^a$ matrices.  For the 3-gluon vertex, this is done through
\leqn{changef}, or, by the use of \leqn{Ttrace}, through
\beq
  -gf^{abc} = {ig\over \sqrt{2}} \tr [T^a T^b T^c - T^a T^c T^b] \ . 
\eeq{decthreeg}
For the 4-gluon vertex, we need to apply this decomposition twice.
The textbook form of the 4-gluon vertex is shown
in Fig.~\ref{fig:fourglue}.  Each term can be manipulated as follows
\beqa
 -ig^2 f^{abe}f^{cde} &=& i {g^2\over 2} \tr([T^a,T^b] [T^c,T^d])\CR
  &=&  i {g^2\over 2} \tr( T^a T^b T^c T^d - T^a T^b T^d T^c - 
     T^b T^a T^c T^d + T^b T^a T^d T^c )
\eeqa{Tchain}
The full 4-gluon vertex can then be rearranged into 
\beq
i {g^2 \over 2 } \tr (T^aT^bT^cT^d) [ 2 g^{\mu\lambda} g^{\nu\sigma} 
     -   g^{\mu\nu} g^{\lambda\sigma} -  g^{\mu\sigma} g^{\nu\lambda}]
\eeq{decomfourg}
plus 5 more terms corresponding to the other 5 color structures in 
\leqn{ggggcolorDec}.  

\begin{figure}
\begin{center}
\includegraphics[height=2.5in]{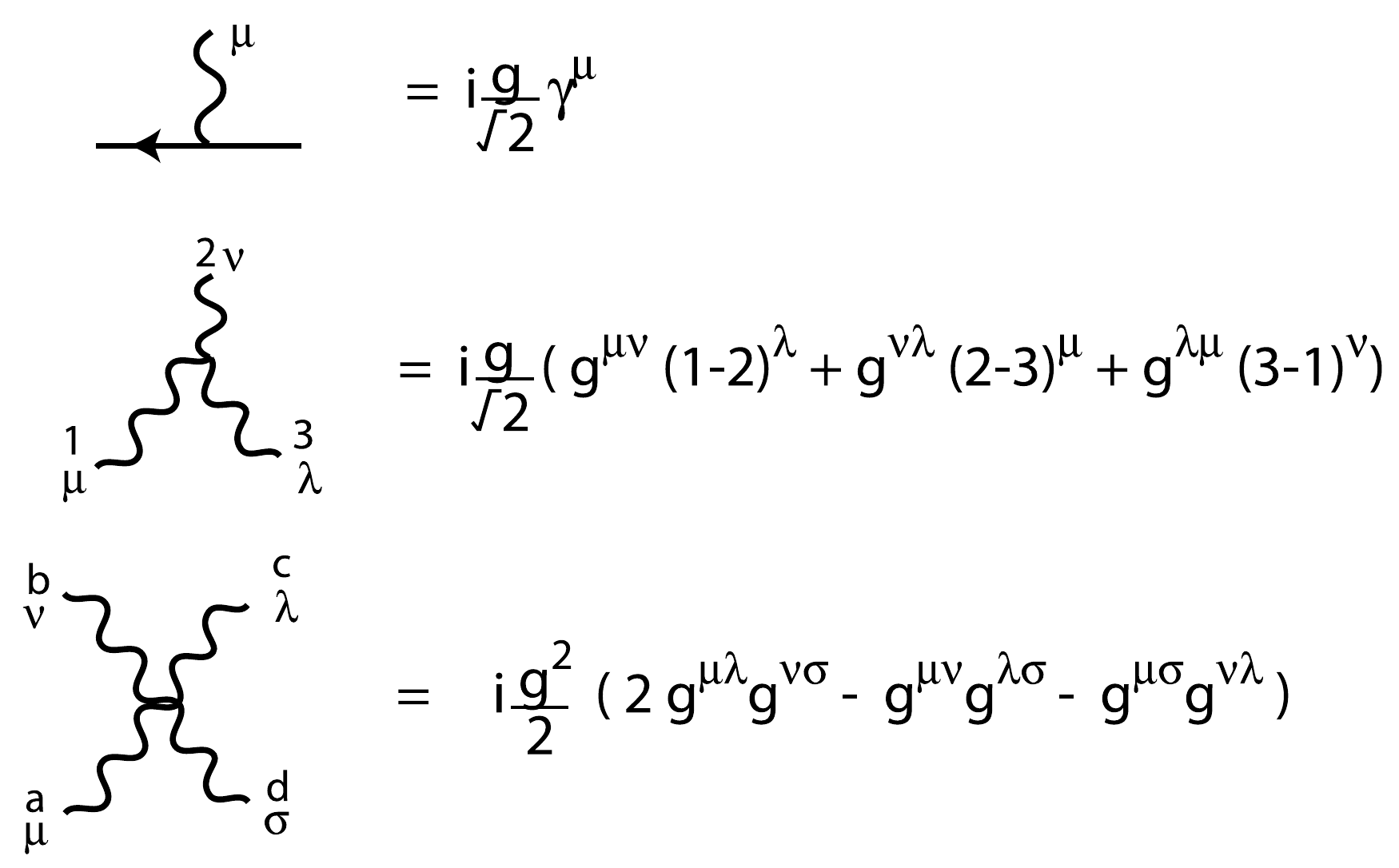}
\caption{Color-ordered Feynman rules for QCD.}
\label{fig:coFeynman}
\end{center}
\end{figure}

These vertices in the form of traces over $T^a$'s can be contracted using
the identity for $SU(N)$ generators
\beq
   T^a_{ij} T^a_{k\ell} = \delta_{i\ell} \delta_{kj} - {1\over N} 
                      \delta_{ij} \delta_{k\ell} \ .
\eeq{TTcontract}
The coefficients in this equation are determined by the normalization
\leqn{Ttrace}, with  $\delta^{aa} = N^2-1$, the number of generators of 
$SU(N)$, and by the requirement that $\tr T^a = 0$.  Using this identity,
a product of traces can be transformed into a single trace plus smaller
factors, for example,
\beq
    \tr [ T^a A] \tr [ T^a B]  =   \tr AB - {1\over N} \tr[A] \tr[B] \ .
\eeq{combinetraces}
Using these methods, the value of a Feynman diagram is naturally organized
as a sum of color-ordered terms.  The individual color-ordered amplitudes
are computed with {\it color-ordered Feynman rules}, shown in 
Fig.~\ref{fig:coFeynman}.  

\begin{figure}
\begin{center}
\includegraphics[height=1.2in]{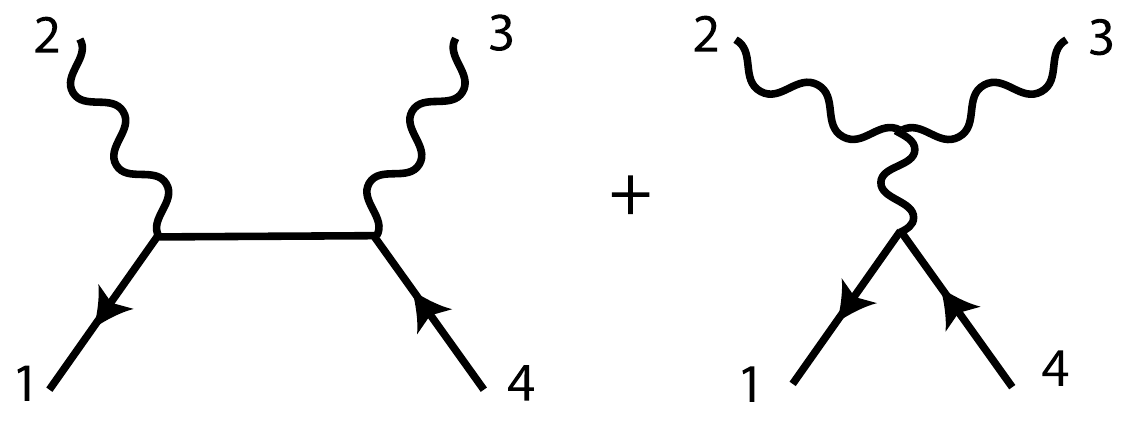}
\caption{Feynman diagrams contributing to the first color-ordered amplitude
 in $q \bar q\to gg$.}
\label{fig:qqggco}
\end{center}
\end{figure}

Using these rules applied to the two diagrams in Fig.~\ref{fig:qqggco},
it is easy to rederive the color-ordered amplitude \leqn{colororderedqqgg}.

\begin{figure}
\begin{center}
\includegraphics[height=1.2in]{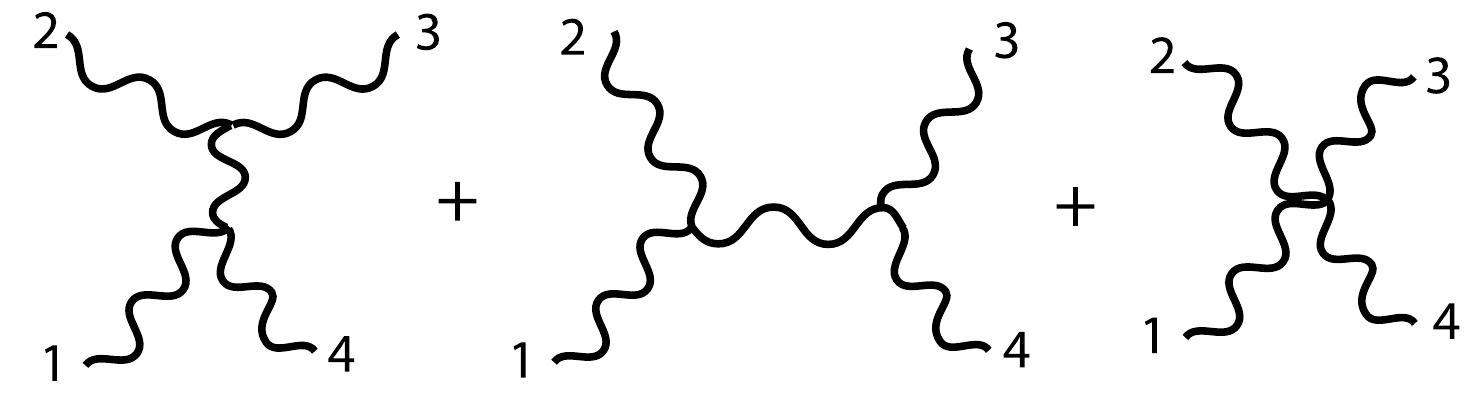}
\caption{Feynman diagrams contributing to the first color-ordered amplitude
 in $gg\to gg$.}
\label{fig:ggggco}
\end{center}
\end{figure}

As an example, I will compute the color-ordered amplitude $\bfM(1234)$
that is used to build up the four-gluon amplitude.  Of the four
Feynman diagrams in Fig.~\ref{fig:gggg}, only the three diagrams shown
in Fig.~\ref{fig:ggggco} contribute to this color-ordered component. 
Here and in the rest of the lectures, the color-ordered amplitudes
in the figures will be ordered clockwise. Using
the Feynman rules in Fig.~\ref{fig:coFeynman}, we find
\beqa
 i\bfM(1234) &=& ({ig\over \sqrt{2}})^2 \biggl[ 
       {-i\over s_{14}}[\eps(4)\cdot \eps(1) (4-1)^\lambda + \eps^\lambda(1)
  ( 2 \, 1 + 4) \cdot \eps(4) + \eps^\lambda(4) (-2\, 4 - 1)\cdot \eps(1)]\CR
    & & \hskip 0.3in \cdot[\eps(2)\cdot \eps(3) (2-3)_\lambda + \eps_\lambda(3)
    ( 2 \, 3 + 2) \cdot \eps(2) + \eps_\lambda(2) (-2\, 2 - 3)\cdot \eps(3)]\CR
& & + {-i\over s_{12}}[\eps(1)\cdot \eps(2) (1-2)^\lambda + \eps^\lambda(2)
    ( 2 \, 2 + 1) \cdot \eps(1) + \eps^\lambda(1) (-2\, 1 - 2)\cdot \eps(2)]\CR
    & & \hskip 0.3in \cdot[\eps(3)\cdot \eps(4) (3-4)_\lambda + \eps_\lambda(4)
  ( 2 \, 4 + 3) \cdot \eps(3) + \eps_\lambda(3) (-2\, 3 - 4)\cdot \eps(2)]\CR
& &\hskip -0.3in  + (-i)[ 2 \eps(1)\cdot \eps(3)\, \eps(2)\cdot \eps(4)
   - 
\eps(1)\cdot \eps(2)\, \eps(3)\cdot \eps(4) - 
\eps(1)\cdot \eps(4)\, \eps(2)\cdot \eps(3) ] \biggr] \ .
\eeqa{bfMforgggg}
With some effort, you can show that this expression obeys the Ward identity.
That demonstrates explicitly that this color component is independently 
gauge-invariant, as required.

\section{MHV amplitudes}

Now that we have reduced the $q \bar q gg$ and $gggg$ QCD amplitudes to 
managable components, it is time to evalute these expressions. We will
see that the values we find fall into surprisingly simple forms.

\subsection{$q\bar q  gg$  amplitude}

 Begin with 
the $q\bar q gg$ amplitude \leqn{colororderedqqgg}.  As in the example
with $\ee\gamma\gamma$ in Section~2, we need to consider all possible
cases of gluon helicity.  From here on, I will notate helicity states by
$+$, $-$ instead of $R$, $L$.

Consider first the case $g_+g_+$.  Choose the 
polarization vectors with reference vector $r = 1$,
\beq
  \eps^\mu(2) = {1\over \sqrt{2}}{\apb{ 1 {\gamma^\mu} 2} \over \spa 12} \qquad
  \eps^\mu(3) = {1\over \sqrt{2}}{\apb{1 {\gamma^\mu} 3} \over \spa 13} \ .
\eeq{pluspluseps}
As with \leqn{smashMeegg}, the first line of \leqn{colororderedqqgg} is
proportional to $\spa 11 = 0$.  Also, in the second line of  
\leqn{colororderedqqgg}, each term contains one of the elements
 $\eps(2)\cdot \eps(3)$,
 $\langle 1 \gamma\cdot \eps(2) 4]$, and $\langle 1 \gamma\cdot \eps(3) 4]$,
all of which are proportional to $\spa 11$.  So, the entire expression
vanishes.  A similar argument shows that the color-ordered amplitude
with two negative helicities vanishes.

The two cases with one positive and one negative gluon helicity are 
distinct color-ordered
amplitudes, not related by Bose symmetry.  So, both must be
computed.  I will begin with the case $g_+(2) g_-(3)$.  Choose
for the polarization vectors
\beq
  \eps^\mu(2) = {1\over \sqrt{2}}{\apb{ 1 {\gamma^\mu} 2} \over \spa 12} \qquad
  \eps^\mu(3) = -{1\over \sqrt{2}}{\bpa{ 4 {\gamma^\mu} 3} \over \spb 43} \ .
\eeq{plusminuseps}
Then all terms in $\bfM$ vanish except for the term in the second line
with $\eps(2)\cdot \eps(3)$.  This factor is
\beq
  \eps(2) \cdot \eps(3) = - {1\over 2} { 2 \spa 13 \spb 42\over \spa 12
\spb43 }  \ .
\eeq{epsproduct}
Then
\beqa
  i\bfM &=& {ig^2\over 2}{ \apb{ 1\, {(2-3)}\, 4} \over \spa 23 \spb 32}
          \bigl(   - {\spa 13 \spb 42\over \spa 12 \spb 43} \bigr)\CR
&=&  -ig^2  { {\spa 13}^2 \spb 42 \over \spa 23 \spb 32 \spa 12}
\eeqa{almostfinishqqgg}
Again, we can find a simpler form by multiplying top and bottom by $\spa 13$.
Using $\spa 13 \spb 32 = - \spa 14 \spb 42$, we find
\beq
  i\bfM = {ig^2}{ {\spa 13}^3 \spa 43 \over \spa 12 \spa 23 \spa 34
    \spa 41 } \ .
\eeq{findqqggpm}

For the case  $g_-(2) g_+(3)$, choose
for the polarization vectors
\beq
  \eps^\mu(2) = -{1\over \sqrt{2}}{\bpa {3 {\gamma^\mu} 2}\over \spb 32} \qquad
  \eps^\mu(3) = {1\over \sqrt{2}}{\apb{ 2 {\gamma^\mu} 3} \over \spa 23} \ .
\eeq{minuspluseps}
For this choice of reference vectors,
\beq 
  \eps(2)\cdot \eps(3) = 0\ , \qquad   2\cdot \eps(3) = 3 \cdot \eps(2) = 0\ .
\eeq{epsvanishes}
Then only the first line of  \leqn{colororderedqqgg} is nonzero.  The value
of this line is
\beqa
    i\bfM &=& {-ig^2\over 2} (-\half) {2^2\over \spa 23 \spb 32}
    { \spa 12 [ 3 (1+2) 2 \rangle \spb 34 \over \spa 12 \spb 21 }\CR
  &=&  ig^2 { \spa 12 \spb 31 \spb 34 \over \spa 23 \spb 32 \spb 21 }  \ .
\eeqa{firstformp}
Now multiply top and bottom by $\spa 12 \spa 42$.  After some rearrangements,
all of the square brackets cancel, and we find
\beq
  i\bfM = {ig^2}{ {\spa 12}^3 \spa 42 \over \spa 12 \spa 23 \spa 34
    \spa 41 } \ .
\eeq{findqqggmp}

These results are very interesting.  The $q\bar q gg$ amplitude vanishes 
when all of the gluon helicities are identical. This followed in a very 
straightforward way from the choice of reference vectors in 
\leqn{pluspluseps}.  It is not hard to see that this method extends to 
prove the vanishing of the 
 amplitude for $q\bar q$ plus any number of positive helicity gluons.
Thus, for any number of gluons, the first nonvanishing tree amplitudes are 
those with one negative helicity gluon and all other gluon helicities positive.
These amplitudes are called the {\it maximally helicity violating} or MHV
amplitudes. In the above examples, these amplitudes are built only from 
angle brackets, with no square brackets, and have the form
\beq
  i\bfM(q_-(1) g_+(2) \cdots g_-(i) \cdots g_+(n-1) \bar q_+(n)) = 
 {ig^{n-2}}{ {\spa 1i}^3 \spa ni \over \spa 12 \spa 23 \cdots 
      \spa {(n-1)} n \spa n1} \ ,
\eeq{MHVqqbar}
where $i$ denotes the gluon with negative helicity.  This formula is in fact
correct for all $i$, $n$.  The complex conjugates of these amplitudes give
the amplitudes for the case of one positive helicity gluon and all other
gluon helicities negative,
\beq
  i\bfM(q_-(1) g_-(2) \cdots g_+(i) \cdots g_-(n-1) \bar q_+(n)) = 
(-1)^{n-1} {ig^{n-2}}{{\spb 1i} {\spb ni}^3 \over \spb 12 \spb 23\cdots 
      \spb {(n-1)} n \spb n1} \ .
\eeq{MHVqqbarconj}
I will give a proof of these formulae in Section 6.

\subsection{Four-gluon  amplitude}

A very similar analysis can be applied to the four-gluon amplitude.  Consider 
first the case with all positive helicities.  Choose the gluon polarization 
vectors so that the same reference vector $r$ is used in every case,
\beq
    \eps^\mu(j) =  {1\over \sqrt{2}} {\apb{r {\gamma^\mu} j} \over \spa rj} \ .
\eeq{allr}
Then, for all $i,j$, 
\beq
    \eps(i)\cdot \eps(j) \sim   \spa rr \spb ji = 0 \ .
\eeq{epsepszero}
By inspection of \leqn{bfMforgggg}, every term contains at least one 
factor of $\eps(i)\cdot \eps(j)$.  Thus, the entire expression vanishes.

This argument is easily extended to the case with one negative helicity
gluon.  The amplitude \leqn{bfMforgggg} is cyclically symmetric, so we can 
chose the gluon 1 to have negative helicity without loss of generality.
Then let
\beq
   \eps^\mu(1) = -{1\over \sqrt{2}} {\bpa {2 {\gamma^\mu} 1} \over \spb 21} \ ,
\qquad 
    \eps^\mu(j) =  {1\over \sqrt{2}} {\apb {1 {\gamma^\mu} j} \over \spa 1j}\ ,
\eeq{epschoicemppp}
for $j = 2,3,4$.  Again, $\eps(i)\cdot \eps(j) = 0$ for all $i,j$, and so
the complete amplitude vanishes.

It is not difficult to see that these arguments carry over directly to 
the $n$-gluon color-ordered amplitudes for any value of $n$.  The tree 
amplitudes
with all positive helicities, or with one negative helicity and all of the 
rest positive, vanish.  The {\it maximally helicity violating} amplitudes
are those with two negative and the rest positive helicities.

It is worth noting that the vanishing of the amplitudes with zero or one
 negative
helicity, both for the $q\bar q + n$ gluon case and for the pure gluon 
amplitudes, is related to supersymmetry.  QCD is not a supersymmetric
theory, but it is an orbifold reduction of supersymmetric QCD.  In 
supersymmetric QCD, these scattering amplitudes are related by supersymmetry
to amplitudes with four external fermions, which must contain two negative
helicities by helicity conservation along the two fermions lines.
A precise discussion of these points can be found in \cite{ManganoParke}.
Note that these arguments apply only to tree amplitudes. The forbidden
amplitudes become
 nonzero at one loop in a nonsupersymmetric theory.

For the 4-gluon amplitude, all that remains is to compute the color-ordered
amplitude in the case with two negative helicities.  By the cyclic 
invariance of $\bfM$, there are only two cases, that in which the two 
negative helicities are adjacent and that in which they are opposite.  As 
an example of the first case, we can analyze $i\bfM(1_-2_-3_+4_+)$.  Choose
the polarization vectors to be
\beqa
   \eps^\mu(1) = -{1\over \sqrt{2}} {\bpa { 4 {\gamma^\mu} 1}\over \spb 41} 
&\qquad& \eps^\mu(2) = -{1\over \sqrt{2}}{\bpa{  4 {\gamma^\mu} 2}
   \over \spb 42}
\CR
   \eps^\mu(3) = {1\over \sqrt{2}} {\apb{ 1 {\gamma^\mu} 3} \over \spa 13} 
&\qquad&  \eps^\mu(4) = {1\over \sqrt{2}} {\apb {1 {\gamma^\mu} 4}
  \over \spa 14}
\eeqa{epsformmpp}
With this choice, all scalar products  of $\eps$'s are zero except for
\beq
\eps(2)\cdot \eps(3) = -\half {2 \spb 43 \spa 12\over \spb 42 \spa 13} = 
     - {\spa 12 \spb 43 \over \spa 13 \spb 42}
\eeq{onenonzeroeps}
Looking back at \leqn{bfMforgggg}, we see that the first and third lines
are zero.  In the second line, the only nonzero term is the one that 
involves $\eps(2)\cdot \eps(3)$ and no other dot product of $\eps$'s.  Thus,
\beqa
   i\bfM &=&   ( {ig^2\over 2}) {-i\over s_{34}} (-4) \eps(2)\cdot\eps(3) \,
   2\cdot \eps(1) \, 3 \cdot \eps(4) \CR 
  &=&  -ig^2 { {\spa 12}^2 \spb 34 \over \spa 34 \spb 41 \spa 41}
\eeqa{mmreduce}
Multiplying top and bottom by $\spa 12$ and rearranging terms in the 
denominator to cancel out the square
bracket factors, we find
\beq
  i \bfM(1_-2_-3_+4_+) = ig^2 { {\spa 12}^4 \over \spa 12 \spa 23\spa 34 
      \spa 41}   \ .  
\eeq{finalmmpp}

Similarly, to evaluate $i\bfM(1_-2_+3_-4_+)$, choose the polarization vectors
to be 
\beqa
   \eps^\mu(1) = -{1\over \sqrt{2}} {\bpa {4 {\gamma^\mu} 1}\over \spb 41} 
&\qquad&  \eps^\mu(2) = {1\over \sqrt{2}} {\apb{ 1 {\gamma^\mu} 2}
  \over \spa 12}
\CR
   \eps^\mu(3) = -{1\over \sqrt{2}} {\bpa{4 {\gamma^\mu} 3}\over \spb 43} 
&\qquad&  \eps^\mu(4) = {1\over \sqrt{2}} {\apb {1 {\gamma^\mu} 4}
   \over \spa 14}
\eeqa{epsformpmp}
With this choice, all scalar products  of $\eps$'s are zero except for
\beq
\eps(2)\cdot \eps(3) =
     - {\spa 13 \spb 42 \over \spa 12 \spb 43}
\eeq{onenonzeroepstwo}
Again, only the term that involves $\eps(2)\cdot \eps(3)$ and no other 
dot product of $\eps$'s
is nonzero.  The value of that term is again given by the 
first line of \leqn{mmreduce}, which, in this case, evaluates to
\beq
i\bfM =  -ig^2 { {\spa 13}^2 {\spb 42}^2 \over \spa 34 \spb 41 \spa 41 
        \spb 43} \ .
\eeq{reducempmp}
Multiplying top and bottom by ${\spa 13}^2$ and rearranging terms in the 
denominator to cancel out the square brackets, we find
\beq
  i \bfM(1_-2_+3_-4_+) = ig^2 { {\spa 13}^4 \over \spa 12 \spa 23\spa 34 
      \spa 41} \ .    
\eeq{finalmpmp}

The form of \leqn{finalmmpp} and \leqn{finalmpmp} strongly suggests that the 
general form of an $n$-gluon MHV amplitude is
\beq
 i\bfM(g_+(1) \cdots g_-(i) \cdots g_-(j) \cdots g_+(n)) = 
     ig^{n-2}{ {\spa ij}^4 \over \spa 12 \spa 23 \cdots 
      \spa {(n-1)} n \spa n1} \ .
\eeq{genMHVg}
The corresponding formula, exchanging positive and negative helicities, is
\beq
 i\bfM(g_-(1) \cdots g_+(i) \cdots g_+(j) \cdots g_-(n)) = 
  (-1)^{n} ig^{n-2}{ {\spb ij}^4 \over \spb 12 \spb 23 \cdots 
      \spb {(n-1)} n \spb n1} \ .
\eeq{genMHVgbar}
I will give a proof of these formulae for all $i,j,n$ in Section 6.  The 
formula \leqn{genMHVg} was discovered in 1986 by Parke and 
Taylor~\cite{PT}.  This 
was the original breakthrough that gave the impetus for all of the work
discussed in the latter sections of this review.

\section{Parton-parton scattering}

Before going deeper into the theory of the MHV formulae presented in the 
previous section, I will present a simple application of these formulae.
The basic ingredient for collider physics is the set of tree-level
cross sections for parton-parton scattering.  These are straightforward to
derive from the QCD Feynman rules, and yet the calculations can be tedious
for students.  In a one-year course in quantum field theory, this subject
generally arises just at that point in the year when the professor's 
family needs a weeklong ski vacation.  Thus,
 in the textbooks, the derivation of
these formulae is typically left as an exercise for the students
 without detailed explanation.
  In this section, I will
show that the MHV formulae make these derivations trivial.

\begin{figure}
\begin{center}
\includegraphics[height=1.2in]{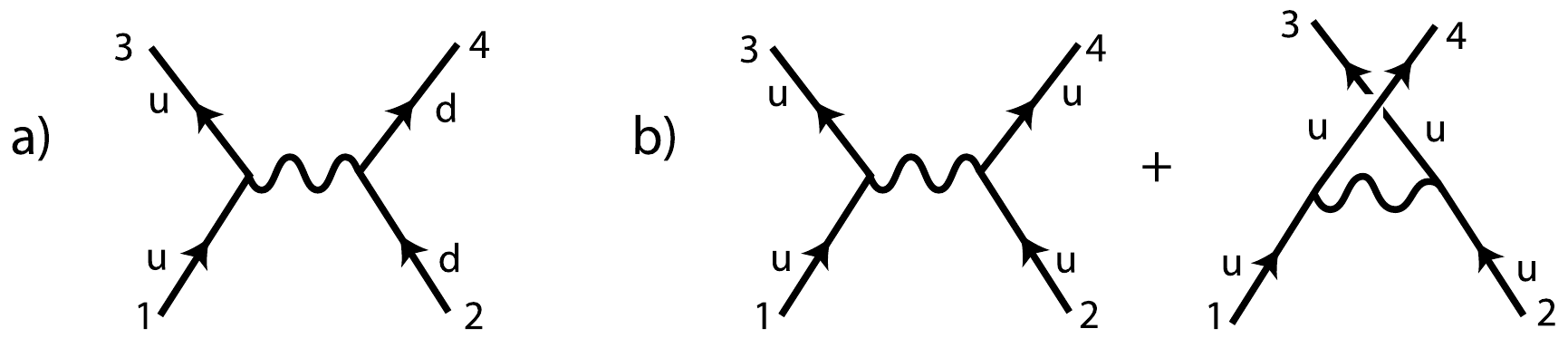}
\caption{Feynman diagrams for quark-quark scattering: (a) $ud \to ud$, 
(b) $uu\to uu$.}
\label{fig:udud}
\end{center}
\end{figure}

\subsection{Four-fermion processes}

Begin with quark-quark scattering.  For scattering of quarks of different 
flavor, there is only one Feynman diagram, shown in Fig.~\ref{fig:udud}(a).
The value of this diagram can be found from \leqn{allangle}.  In this formula,
all momenta are outgoing, so we simply need to cross the correct particles
into the initial state.  Which particles should be crossed depends on which
helicity amplitude we are considering, for example, 
$u_L d_L \to u_L d_L$ or $u_L d_R \to u_L d_R$. We also need to include
the QCD color matrices.  With the numbering of lines as in the figure,
with the arrows indicating the direction of {\it left-handed} fermions,
the matrix elements are
\beqa
   i \M(u_L d_L \to u_L d_L) &=& ig^2{ {\spa 34}^2 \over \spa 13 \spa 24}
         T^a_{31} T^a_{42} \CR
   i \M(u_L d_R \to u_L d_R) &=& -ig^2{ {\spa 23}^2 \over \spa 13 \spa 24}
         T^a_{31} T^a_{42} \CR
\eeqa{udscatter}
The scattering of identical quarks, for example, $u_Lu_L \to u_L u_L$, 
is given by two Feynman diagrams, shown in Fig.~\ref{fig:udud}(b).
For this case
\beq
  i  \M(u_L u_L \to u_L u_L) = ig^2\biggl[ { {\spa 34}^2 \over \spa 13 \spa 24}
         T^a_{31} T^a_{42}  -{ {\spa 43}^2 \over \spa 14 \spa 23}
         T^a_{41} T^a_{32}\biggr]  \ .
\eeq{Muu}
The extra minus sign in the second term comes from interchange of 
identical fermions.  To compute cross sections, we square these expressions,
sum over final
helicities and colors, and average over initial helicities and colors.  
The interference term in the square of \leqn{Muu} requires
\beq
    \spa 13 \spa 24 (\spa 14 \spa 23)^* = \spa 13 \spb 32 \spa 24 \spb 41 = 
   - \spa 13 \spb 31 \spa 14 \spb 41 = - tu \ .
\eeq{interferencedenom}
By parity, the cross sections are unchanged when all helicities are 
reversed. The cross 
sections for antiquarks can be computed by applying crossing symmetry,
so the formulae above are all that we need to cover all of the 
possible cases.

For the color sums and averages, here and in the later calculations in
this section, we will need the formula 
$T^aT^a = {8\over 3}{\bf 1}$
and the traces
\beq
  \tr[ T^aT^aT^bT^b] = {64\over 3} \ , \quad
  \tr[ T^aT^bT^aT^b] = - {8\over 3} \ .
\eeq{colorsums}
These are easily proved using \leqn{TTcontract}. The second trace appears
in the interference term in the square of \leqn{Muu}. 
For a general $SU(N)$ gauge
group, these results are
\beq
  \tr[ T^aT^aT^bT^b] = {(N^2-1)^2\over N} \ , \quad
  \tr[ T^aT^bT^aT^b] = - {(N^2-1)\over N} \ ,
\eeq{colorsumsforN}
so that the second color sum is suppressed by $1/N^2$.  This is an
example of the general result that interference terms between different color
structure are suppressed by $1/N^2$.  That result
follows from the broad picture
of $1/N$ counting presented by 't Hooft in~\cite{tHooft}.

Assembling the pieces, we find
\beqa
    {d\sigma\over d\cos\theta}(ud\to ud) &=& {2\over 9} {\pi \alpha_s^2\over s}
 \biggl[ {s^2 + u^2\over t^2} \biggr] \CR
    {d\sigma\over d\cos\theta}(uu\to uu) &=& {2\over 9} {\pi \alpha_s^2\over s}
 \biggl[ {s^2 + u^2\over t^2} + {s^2 + t^2 \over u^2}
      - {2\over 3}  {s^2\over tu}\biggr]  \ ,
\eeqa{qqscatter}
where $\theta$ is the scattering angle in the parton-parton center of 
mass frame.
The second of these formulae is to be integrated over $cos\theta > 0$ only.
Crossing these formulae into other channels,
\beqa
    {d\sigma\over d\cos\theta}(u\bar d\to u\bar d) 
&=& {2\over 9} {\pi \alpha_s^2\over s}
 \biggl[ {s^2 + u^2\over t^2} \biggr] \CR
    {d\sigma\over d\cos\theta}(u\bar u\to u\bar u) 
&=& {2\over 9} {\pi \alpha_s^2\over s}
 \biggl[ {s^2 + u^2\over t^2} + {t^2 + u^2 \over s^2}
      - {2\over 3}  {u^2\over st}\biggr]  \ .
\eeqa{qqbarscatter}

\begin{figure}
\begin{center}
\includegraphics[height=1.5in]{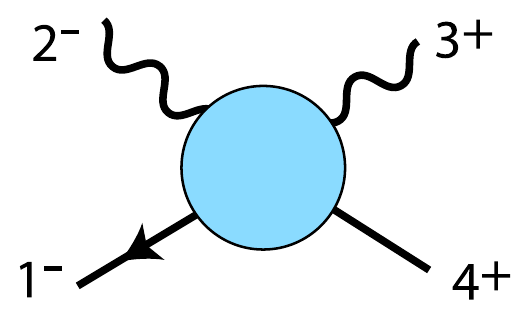}
\caption{Notation for the process $u_L\bar u_R \to g_L g_R$.}
\label{fig:qqggproc}
\end{center}
\end{figure}

\subsection{$q\bar q gg$  processes}

Next, consider $q\bar q \to gg$. The $q$ and $\bar q$ necessarily have 
opposite helicity.  The gluons also have opposite helicity.  The two cases
of gluon helicity are related by gluon interchange or $t\leftrightarrow u$.
So it suffices to consider the case shown in Fig.~\ref{fig:qqggproc},
$u_L(4)\bar u_R(1) \to g_L(2) g_R(3)$.  For this case, the color-ordered
amplitudes are MHV amplitudes, and so our results from Section 4 give
\beq
  i\M = ig^2 \biggl[  { {\spa 13}^3 \spa 43\over \spa 12 \spa 23 \spa 34
 \spa 41 }  T^a T^b + { {\spa 13}^3 \spa 43\over \spa 13 \spa 32 \spa 24
 \spa 41 }  T^b T^a  \biggr] \ .
\eeq{Muugg}
The square of this amplitude, traced over color, is
\beqa
\tr [ \M |^2 &=&  g^4\biggl( {64\over 3}[ {t u\over s^2} + {t^3\over us^2}]
    - 2 {8\over 3} {t^2\over s^2} \biggl) \CR
  &=&  {64\over 3}g^4 [ { t\over u} - {9\over 4} {t^2 \over s^2} ] \ .
\eeqa{Muuggeval}
From the first line to the second, I have used $u^2 + t^2 = (u+t)^2 - 2 ut = 
s^2 - 2 ut$.  Adding the case with the two gluons interchanged, and including
the factors for averaging over initial colors and helicities, we find 
as the final result,
\beq
    {d\sigma\over d\cos\theta}(u\bar u\to gg) = {16\over 27} 
 {\pi \alpha_s^2\over s}
 \biggl[ {t\over u} + {u\over t} - {9\over 4}{t^2+u^2\over s^2} \biggr] \ .
\eeq{uuggfinal}
This formula should be integrated over $\cos\theta > 0$ only.  Crossing
this formula into other channels gives
\beqa
    {d\sigma\over d\cos\theta}(u g\to ug) &=& {2\over 9} 
 {\pi \alpha_s^2\over s}
 \biggl[ -{s\over u} - {u\over s} + {9\over 4}{s^2+u^2\over t^2} \biggr] \CR
    {d\sigma\over d\cos\theta}(gg\to u\bar u) &=& {1\over 12} 
 {\pi \alpha_s^2\over s}
 \biggl[ {t\over u} + {u\over t} - {9\over 4}{t^2+u^2\over s^2} \biggr] \ .
\eeqa{gguufinal}

\subsection{Four-gluon  processes}

Finally, we need to compute the $gg\to gg$ scattering amplitude.  This can be
done directly from the formalism we have already developed, but it 
is useful to add one additional trick.  

Extend the gauge group 
from $SU(N)$ to $U(N)$ by adding an 
extra generator $T^0 = {{\bf 1}/\sqrt{N}}$. Since the 3- and 4-gluon vertices
are proportional to the structure constants $f^{abc}$ and the extra
$U(1)$  generator commutes with the $SU(N)$
generators, the additional boson gives no contribution to the cross 
sections. However, adding this term simplifies the color algebra.
The contraction identity \leqn{TTcontract} now takes the simpler form
\beq
   T^a_{ij} T^a_{k\ell} = \delta_{i\ell} \delta_{kj} \ .
\eeq{TTcontractU}
We can use this identity to compute the square of the 4-gluon 
amplitude \leqn{ggggcolorDec}.  The squares of the individual color 
traces are equal to
\beq
   \tr [ T^a T^b T^c T^d ] \, \tr [T^d T^c T^b T^a] = N^4 = 81 \ .
\eeq{directtrace}
The cross terms are proportional to 
\beq
   \tr [ T^a T^b T^c T^d ] \, \tr [T^d T^c T^a T^b] = 
  \tr [ T^a T^b T^c T^d ] \, \tr [T^d T^a T^b T^c] = N^2 = 9 \ .
\eeq{crosstrace}
Then, letting $I,J = 1,\ldots,6$ index the six possible cyclic orderings of the
gluons, we find
\beqa
  \tr[ |\M|^2 ] &=&  81 \sum_I |\bfM(I)|^2 + 9 \sum_{I\neq J} \bfM(I) \bfM(J)^*
   \CR
 &=&  72 \sum_I |\bfM(I)|^2 + 9| \sum_I \bfM(I)|^2 \ .
\eeqa{gcolorsum}
 
\begin{figure}
\begin{center}
\includegraphics[height=1.2in]{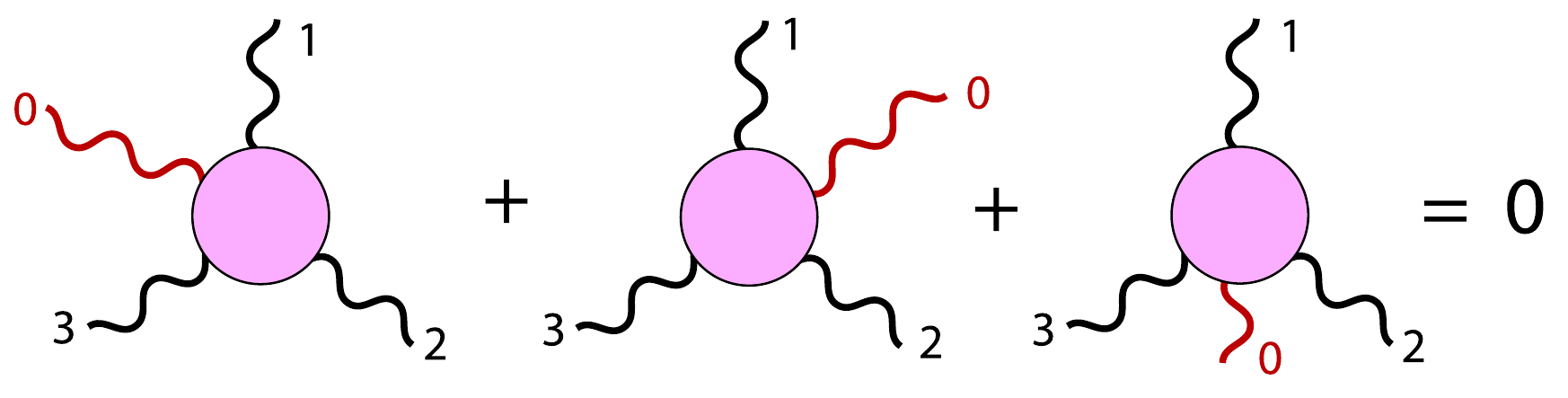}
\caption{The $U(1)$ Ward identity for color-ordered amplitudes.}
\label{fig:UoneWard}
\end{center}
\end{figure}

I explained above that,
 when we add the extra generator to extend $SU(N)$ to $U(N)$, the 
value of $\tr |\M|^2 $ cannot change.  
The vanishing of the coupling to the $T^0$ gauge bosons
is expressed in the color-ordered amplitudes as the Ward
identity shown in Fig.~\ref{fig:UoneWard}:  The sum over all possible 
orderings of a $U(1)$ boson in a  color-ordered $n$-gluon 
amplitudes must vanish. An example of this identity, for 4-gluon 
amplitudes, is
\beq
  i\bfM(0_-1_-2_+3_+) +    i\bfM(1_-0_-2_+3_+) +    i\bfM(1_-2_+0_-3_+) = 0
\eeq{exampleUoneWard}   
We can easily check this from the explicit expressions for MHV amplitudes.
Evaluating the left-hand side of \leqn{exampleUoneWard} gives
\beqa 
& & ig^2{ \spa 01}^4 \biggl[ {1\over \spa 01 \spa 12 \spa 23 \spa 30}
  + {1\over \spa 10 \spa 02 \spa 23 \spa 31} + 
 {1\over \spa 12 \spa 20 \spa 03 \spa 31} \biggr] \CR
& &  = ig^2 {\spa 01}^4 {\bigl[
  \spa 02 \spa13 + \spa 03 \spa 21 + \spa 01 \spa 32\bigr]
 \over \spa 01 \spa 12 \spa 23 \spa 30 \spa 02 \spa 13 }
\eeqa{reduceUone}
 The expression in brackets in the second line
 vanishes  by the Schouten identity 
 \leqn{Schouten}. 

The $U(1)$ Ward identity implies that  
\beq
 \sum_I \bfM(I)= 0 \ .
\eeq{sumUoneWard}
  Then
\leqn{gcolorsum} becomes
\beq
  \tr[ |\M|^2 ] = 72 \sum_I |\bfM(I)|^2  \ .
\eeq{gcolorsumfinal}

Now we must evaluate this expression for all possible choices of the 
external particle helicities.  Of the 16 possibilities, only 6 involve
2 positive and 2 negative helicities.  All other choices give zero. 
For the color-ordering $1234$, the nonzero amplitudes are
\beqa
|\bfM(1_-2_-3_+4_+)|^2 & =& g^4 \biggl| { {\spa 12}^4\over \spa 12 \spa 23
    \spa 34 \spa 41} \biggr|^2 = g^4 {s^2\over t^2}\CR
|\bfM(1_-2_+3_-4_+)|^2 & =& g^4 \biggl| { {\spa 13}^4\over \spa 12 \spa 23
    \spa 34 \spa 41} \biggr|^2 = g^4 {u^4\over s^2t^2}\CR
|\bfM(1_-2_+3_+4_-)|^2 & =& g^4 \biggl| { {\spa 14}^4\over \spa 12 \spa 23
    \spa 34 \spa 41} \biggr|^2 = g^4 {t^2\over s^2 }\ ,
\eeqa{threeMs}
and the parity conjugates of these amplitudes.  The sum over color orderings
sums over these results crossed into all possible channels.  Thus
\beqa
\sum_h \sum_I \tr |\bf M(I)|^2 &=& 4g^4 \cdot [ {s^2 + u^2\over t^2 } + 
  {t^2 + u^2 \over s^2} + {s^2 + t^2 \over u^2 } + {u^4\over t^2 s^2}
  + {t^4\over u^2 s^2} + {s^4 \over t^2 u^2} ] \CR
 &=&  16 g^4 [ 3 - {su\over t^2 } -  {ut\over s^2} - {st\over u^2}] \ .
\eeqa{sumMs}
Including the factors for initial-state color and helicity averaging, we
find
\beq
    {d\sigma\over d\cos\theta}(gg\to gg) = {9\over 4} 
 {\pi \alpha_s^2\over s}
 [ 3 - {su\over t^2 } -  {ut\over s^2} - {st\over u^2}] \ .
\eeq{ggggfinal}
This formula should be integrated over $\cos\theta > 0$ only.  
This completes the calculation of all of the tree-level $2\to2$ parton-parton
scattering cross sections.

\section{Britto-Cachazo-Feng-Witten recursion}

Now I will  develop some methods that will allow us to prove the 
MHV formulae, and also to use these results to compute the more
compex non-MHV amplitudes.
  The general approach will follow the idea of Cachazo,
Svrcek, and Witten~\cite{Witten,CSW} 
that we should consider the tree-level color-ordered
amplitudes as analytic functions of the angle brackets and square 
brackets.  We can then analytically continue scattering amplitudes to 
complex momenta while keeping the property that all external legs are 
on mass shell.  At the end of the calculation, 
to extract physical results, we will specialize to the
values such that  $\spa ij = ( \spb ji)^*$.

\subsection{Three-point scattering amplitudes}

\begin{figure}
\begin{center}
\includegraphics[height=1.2in]{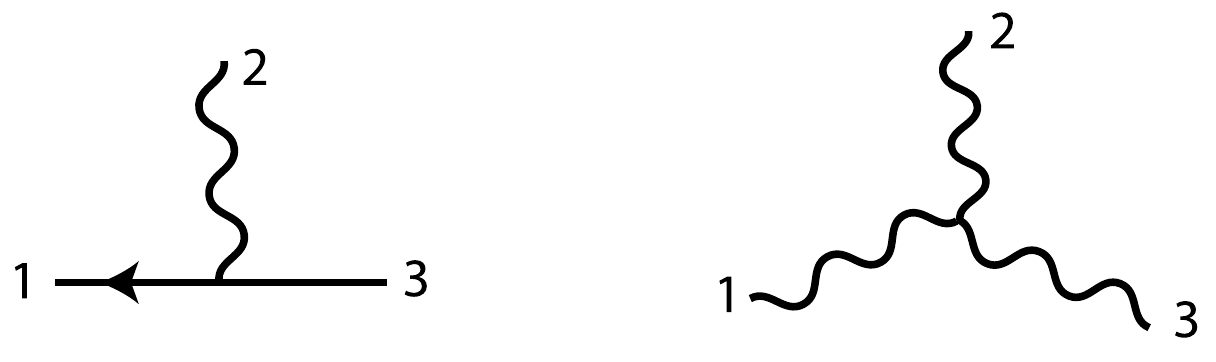}
\caption{Diagrams contributing to the three-point $qg\bar q$ and $ggg$ 
   amplitudes.}
\label{fig:threepoint}
\end{center}
\end{figure}

This freedom to discuss QCD scattering amplitudes for arbitrary complex
momenta allows us to write on-shell three-point scattering amplitudes.
For three-point amplitudes, shown in Fig.~\ref{fig:threepoint}, momentum
conservation implies that $1+2+3 = 0$.  Then
\beq
  0 = 1^2  = (2+3)^2 =   2\, 2\cdot 3 \ ,
\eeq{noinvariant}
and similarly the other two Lorentz products vanish, so it would seem that
there are no invariants for these amplitudes to depend on.  However, 
in terms of square brackets and angle brackets,
\beq
      2\,  2\cdot 3 =  s_{23} = \spa 23 \spb 32 \ .
\eeq{separatesqangle}
Thus, with complex momenta, 
we can satisfy \leqn{noinvariant} by having $\spb 32 = 0$ while keeping
$\spa 23$ nonzero.

With this idea in mind, I will compute the color-ordered amplitude for 
gluon emission from $q\bar q$,
$i\bfM(q_-(1) g_-(2) \bar q_+(3))$.  The amplitude comes from one Feynman
diagram, shown in Fig.~\ref{fig:threepoint}. The value of this diagram
is 
\beq
   {ig\over \sqrt{2}} \apb{ 1 {\gamma^\mu} 3} \, ( - {1\over \sqrt{2}}
      { \bpa {r {\gamma_\mu} 2} \over \spb r2 }) \ ,
\eeq{threepointqgq}
with $r$ an arbitrary lightlike reference vector.  Applying the Fierz
identity, and then multiplying top and bottom by $\spa 12$, this 
rearranges to 
\beq
   -ig {\spa 12 \spb r3 \spa 21 \over \spb r2 \spa 21} = -ig 
       { {\spa 12}^2 \over \spa 31} \ .
\eeq{rearrangethree}
The reference vector $r$ cancels out, and the final result is just the 
MHV amplitude with $n = 3$,
\beq 
  i\bfM(q_-(1) g_-(2) \bar q_+(3)) = ig { {\spa 12}^3 \spa 32 \over
      \spa 12 \spa 23 \spa 31 } \ .
\eeq{finalqgq}

A similar calculation gives the form of the three-gluon vertex.
\beqa
 & & i \bfM(g_-(1) g_-(2) g_+(3))\CR  
  & &  \hskip 0.1in = {ig\over \sqrt{2}}
 [ \eps(1)\cdot \eps(2) (1-2)\cdot \eps(3) \CR
  & & \hskip 0.6in + \eps(2)\cdot \eps(3) (2-3)\cdot
 \eps(1) + \eps(3)\cdot \eps(1) (3-1)\cdot \eps(2)] \ .
\eeqa{threeg}
Choose for the polarization vectors
\beq
  \eps^\mu(1) = - {1\over \sqrt{2}}{ \bpa{ r {\gamma^\mu} 1} \over \spb r1} \ ,
\quad 
  \eps^\mu(2) = - {1\over \sqrt{2}}{ \bpa{ r {\gamma^\mu} 2} \over \spb r2} \ ,
\quad 
  \eps^\mu(3) = {1\over \sqrt{2}}{ \apb{ s {\gamma^\mu} 3 } \over \spa s3} \ , 
\eeq{threeeps}
so that $\eps(1)\cdot \eps(2) = 0$.  Then \leqn{threeg} evaluates to 
\beqa
 & & 
 {ig\over \sqrt{2}} {1\over \sqrt{2}}{ (0 + \spb r3 \spa s2 \bpa {r\,(2-3)\, 1}
     + \spb r3 \spa s1 \bpa {r\, (3-1)\, 2 }) \over
   \spb r1 \spb r2 \spa s3 }\CR
& & = ig {\spb r3 \over \spb r1 \spb r2 \spa s3}  ( \spb r3 \spa 31 \spa 2s
      - \spb r3 \spa 32 \spa 1s) \CR
& & = ig {{\spa 12} {\spb r3}^2 \over \spb r1 \spb r2}
\eeqa{rearrthreeg}
If we multiply top and bottom by ${\spa 12}^2$ and rearrange the denominator,
the factors of $\spa r3$ cancel, and we find
\beq
 i \bfM(g_-(1) g_-(2) g_+(3)) = ig 
      { {\spa 12}^4\over \spa 12 \spa 23 \spa 31}\ .
\eeq{finalggg}

For reference, the corresponding formulae for two positive and one negative
helicity are 
\beq 
 i \bfM(q_-(1) g_+(2) \bar q_+(3)) =  ig { {\spb 12}{\spb 32}^3
      \over \spb 12 \spb 23 \spb 31}\ ,\quad
 i \bfM(g_+(1) g_+(2) g_-(3)) = -ig { {\spb 12}^4
\over \spb 12 \spb 23 \spb 31}\ .
\eeq{conjggg}
I note again that all other $n$-gluon and $q\bar q + n$ gluon amplitudes
with only one negative or positive helicity are zero. Only these cases 
are nonzero, with the MHV values.

\subsection{BCFW recursion formula}

It would be wonderful if we could use these three-point functions as 
building blocks for the construction of general tree amplitudes.  The
usual understanding is that we need {\it off-shell} three-point functions
to build up general amplitudes.  However, this common-sense idea is evaded
by a beautiful strategy of Britto, Cachazo, Feng, and Witten~\cite{BCFW}.

\begin{figure}
\begin{center}
\includegraphics[height=1.5in]{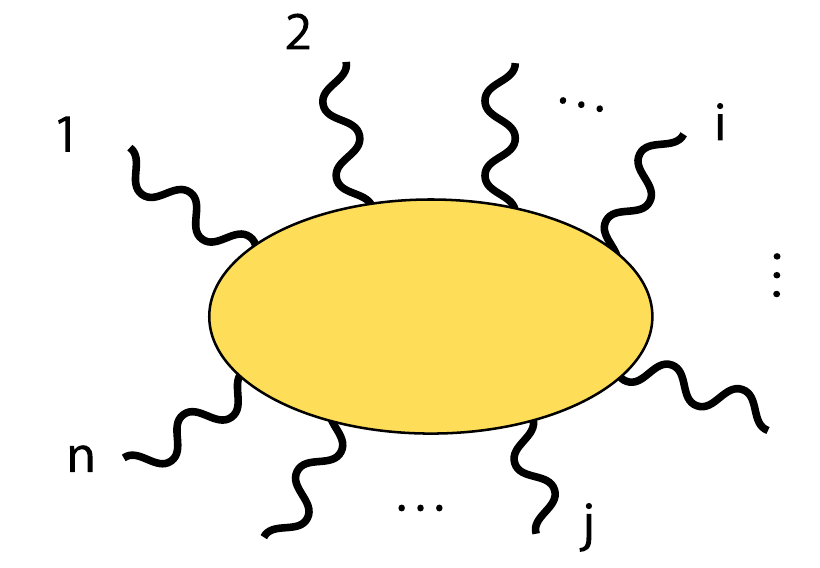}
\caption{Notation for the discussion of the 
      Britto-Cachazo-Feng recursion formula.}
\label{fig:prepBCF}
\end{center}
\end{figure}

Consider a color-ordered amplitude $i\bfM(1 \cdots n)$, as shown 
in Fig.~\ref{fig:prepBCF}.  Choose two legs $i$, $j$, and choose a value of 
$z$, a complex variable.  Now define new momenta $\hat i$ and $\hat j$ by
shifting the square bracket of $i$ and the angle bracket of $j$,
\beqa
     \hat i \rangle = i\rangle \phantom{\ - z i\rangle}
     & \qquad &   \hat i] = i] + z j] \CR
     \hat j \rangle = j\rangle - z i\rangle & \qquad &  \hat j] = j]  \ ,
\eeqa{ijshift}
The  momenta $\hat i$ and $\hat j$ are given by 
\beq
    \hat i = i\rangle [ i + z\  i\rangle [ j \qquad
    \hat j = j\rangle [j - z\  i\rangle [ j  \ .
\eeq{ijvalues}
Notice that $\hat i + \hat j = i+j$, so momentum conservation is respected.
  For general $z$, $\hat i$ 
and $\hat j$ have unphysical complex values.  However, since 
$\hat{i}^2$ is proportional to $\spa ii = 0$, and similarly, 
$\hat{j}^2 = 0$, the new momenta remain on shell.

As a rule of thumb, one should shift the {\it square} bracket of an 
external state with {\it negative} helicity and the {\it angle} 
bracket of an external state with {\it positive} helicity, that is,
$(i,j) = (-,+)$.  This minimizes the number of factors of $z$ in the
numerator.  I will give a more careful analysis of this point below.

Call the amplitude evaluated at the  modified momenta $i\bfM(z)$.
Now consider the integral
\beq
           \oint {dz\over 2\pi i} {1\over z} \, i\bfM(z)
\eeq{BCFintegral}
taken around a large circle in the complex $z$ plane.  If $i\bfM(z) \to 0$
as $z\to \infty$, then this integral vanishes.  In that case, we can set 
to zero the value obtained by contracting the contour and summing over the 
poles that it encloses.  There is  an obvious pole at $z=0$.  The 
residue of this pole is exactly 
\beq
           i\bfM(z = 0) \ ,
\eeq{poleatzero}
the amplitude that we wish to evaluate.  

\begin{figure}
\begin{center}
\includegraphics[height=1.5in]{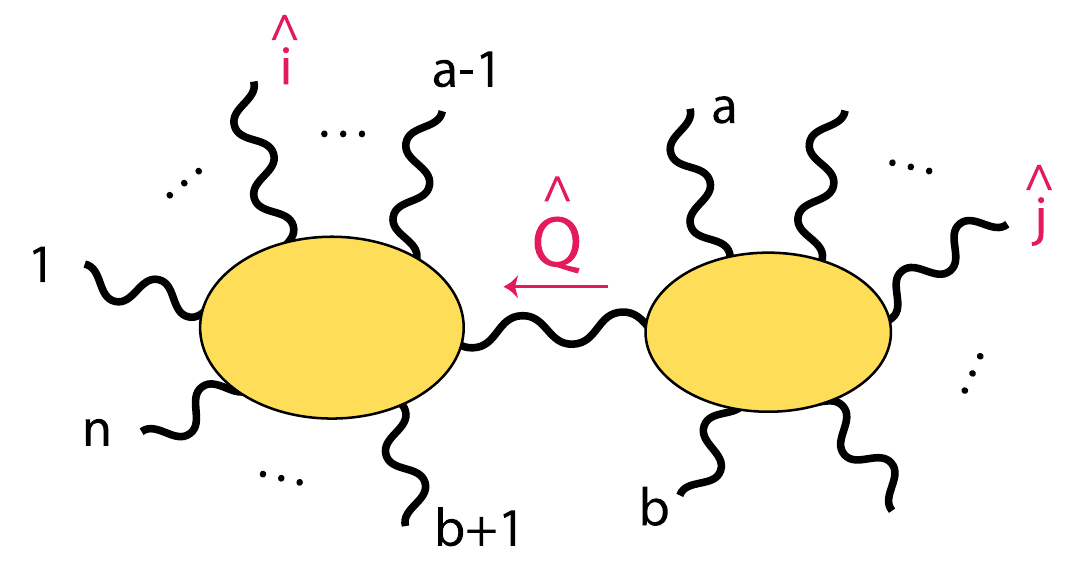}
\caption{A factorization of the  amplitude in Fig.~\ref{fig:prepBCF}
   that gives a pole in the amplitude $i\bfM(z)$.}
\label{fig:BCFWfactorize}
\end{center}
\end{figure}

The other poles come from the 
denominator of the color-ordered amplitude.  We are discussing tree amplitudes,
and so these poles will come from the propagator involved in a factorization
of the amplitude, as shown in Fig.~\ref{fig:BCFWfactorize}.  The denominator 
of this propagator is 
\beq
   \hat{Q}^2 =  s_{a\cdots b}(z) = \bigl( \sum^b_{m = a} k_m \bigr)^2 \ ,
\eeq{saddb}
 the invariant sum of momenta on the 
right, including any terms from the shifts \leqn{ijvalues}.
 If both $i$ and $j$ are on the same side in the factorization,
$s_{a \cdots b}$ is idependent of $z$ and there is no pole.
  If $i$ is on the left side in the factorization and $j$ is on the 
right,
\beqa
        s_{a\cdots b}(z) &=& (\ \sum_{m = a}^b k_m - z i\rangle [ j \ )^2 \CR
                 &=&  s_{a \cdots b}(0) - z \langle i ( \sum_a^b k_m ) j] + 
                       z^2  \spa i i \spb jj  \ .
\eeqa{sabdef}
The last term is zero.  The denominator then has a simple pole at
\beq
          z_* =     {s_{a\cdots b}\over \langle i (\sum_a^b k_m )j] } \ .
\eeq{zstarloc}
The residue of this pole is 
\beqa
& &  {1\over z_*}
     i\bfM((b+1) \ldots \hat i \ldots (a-1) \hat Q)
     {i\over -\langle i(\sum_a^b k_m) j]}
          i\bfM(\hat Q a \ldots  \hat j \ldots b)\CR
   & &   =  -     
    i\bfM((b+1) \ldots \hat i \ldots (a-1) \hat Q) {i\over s_{a \cdots b}}
          i\bfM(\hat Q a \ldots  \hat j \ldots b)
\eeqa{reduceresidue}
The amplitudes in each term are evaluated with the
value $z = z_*$, given by \leqn{zstarloc},  appropriate to the cut that
is chosen.

The result of this analysis is that, if we have chosen the shifted momenta
$i$ and $j$ so that  $i\bfM(z)$ vanishes at infinity,
then 
\beq
  i\bfM(1 \ldots n) = \sum_{a,b} 
     i\bfM((b+1) \ldots \hat i \ldots (a-1) (-\hat Q))\ 
     {i\over s_{a \cdots b}}\
          i\bfM(\hat Q a \ldots \hat j \ldots b)    \  .
\eeq{BCFW}
where $Q = -\sum_a^b k_m$.
This is the {\it BCFW recursion formula}.  An arbitrary color-ordered
helicity amplitude can then be
 evaluated by breaking it down into amplitudes with a
smaller number of legs.  The amplitudes on the right-hand side of \leqn{BCFW}
are evaluated with on-shell but complex-valued momenta.  

The recursion 
must be run until we can immediately evaluate the amplitudes on the 
right-hand side.  Since all nonzero amplitudes with five external legs or
fewer are either MHV or the conjugates of MHV amplitudes, only a few
steps of the recursion are needed in practice.

For QCD with all massless particles, the complete solution of the 
recursion is known.  However, the result is 
not simple, so, in keeping with the philosophy of these lectures, I will
refer you to the reference~\cite{Henn,HennDix,Bourj}  

There is one subtlety that should be clarified to evaluate the right-hand
side of the BCFW recursion formula.  The amplitudes involve angle brackets
and square brackets of the complex momentum $\hat Q$.  In the examples 
given later in our discussion, I will evaluate these brackets by assembling
them into complete factors of the momentum $\hat Q$.  To do this, we will
need to relate the brackets $(-\hat Q)\rangle$ and $(-\hat Q)]$ in the 
amplitude on the left to 
$\hat Q\rangle$
and $\hat Q]$.  It is consistent  always to take
\beq
    (-\hat Q) \rangle = i \hat Q \rangle  \qquad 
    (-\hat Q) ] = i \hat Q ]  \ .   
\eeq{minushats}
One special circumstance should be noted. If the line on which the amplitude
factorizes is a fermion propagator, the value of this propagator is
\beq
         i { Q] \langle Q \over Q^2 }  \quad \mbox \quad i 
              { Q\rangle [ Q \over Q^2 }   \ .
\eeq{fermionprop}
Then  one of the brackets in the left-hand amplitude is a $Q]$, not a 
$(-Q)]$.  To compensate for this, we need to add a factor $(-i)$ for 
a cut through a fermion propagator.

Finally, we need to discuss under what circumstances the amplitude $i\bfM(z)$ 
will vanish at infinity.  I claim that, for the shift given by 
\leqn{ijshift} in which $i$ is shifted in the square bracket and $j$ is 
shifted in the angle bracket, the helicity choices $(i,j) = (-,+), (-,-), 
(+,+)$ give the BCFW recursion formula, while in the case
$(i,j) = (+,-)$ there are  extra
terms from the integral at infinity.  A transparent argument for this 
claim has been given by Arkani-Hamed and Kaplan~\cite{AHK}. 

\begin{figure}
\begin{center}
\includegraphics[height=1.5in]{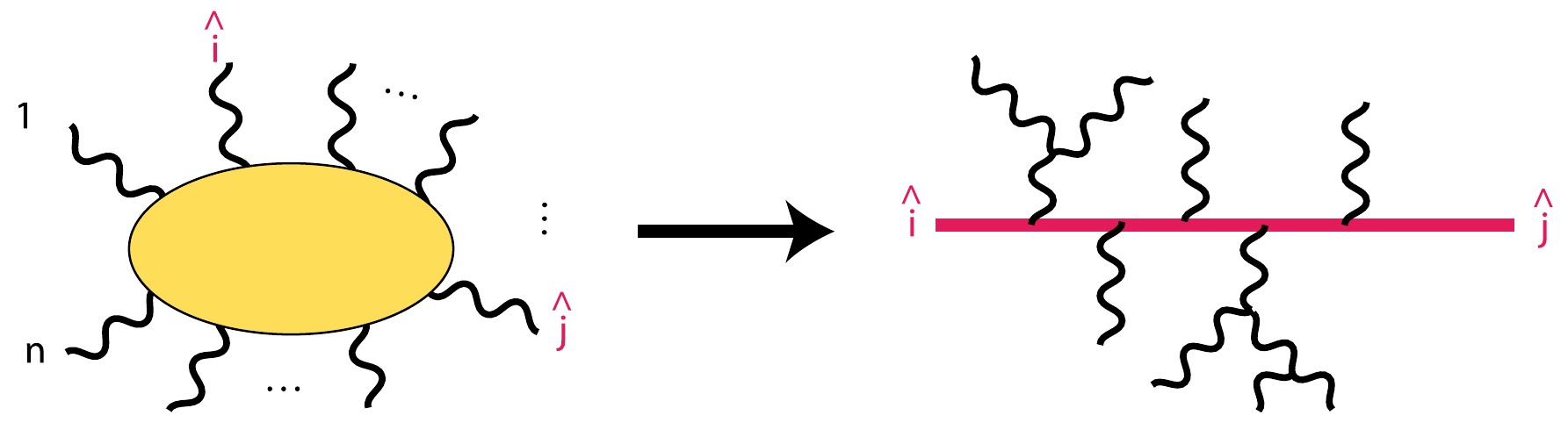}
\caption{Analysis of the $z \to \infty$ behavior of a multigluon tree diagram.}
\label{fig:zlimit}
\end{center}
\end{figure}

In the limit $z\to \infty$, the dominant momentum flowing through the 
diagram is that induced by the shift. Define
\beq 
            q =   i\rangle [j \  \mbox{or} \  q^\mu = \langle i \gamma^\mu j]
\eeq{qAHKdef}
Then the large momentum is $zq$. Note that $q^2 = 0$.
  It is useful to picture the diagrams as
containing one line whose propagators carry  momenta $zq+k$ close to this
large momenta.  This line attaches to propagators that carry momenta of
order 1. This structure is illustrated in Fig.~\ref{fig:zlimit}.
 Each propagator has a denominator of order
\beq
        (zq+k)^2 =  2z q\cdot k + k^2 =  {\cal O}(z) \ .
\eeq{BCFWdenom}
Each vertex is at most linear in momentum, and so is at most of order $z$.
Then each combination of vertex and propagator is of order $z^0$.  The 
chain of vertices and propagators has one extra vertex, so this is of 
order $z$.

Now consider the polarization vectors.  We can choose $i$ or $j$ as the 
reference vector, whichever gives a nonsingular result.
Then
\beqa 
   \eps_+(\hat i) =   {z\langle j \gamma^\mu j] 
     + \langle j \gamma^\mu i]\over \spa ji} + \cdots
        & \qquad &  
   \eps_-(\hat i) =  - {1\over z}{[ i \gamma^\mu i \rangle\over \spb ij} \CR
   \eps_+(\hat j) = - {1\over z}{\langle j \gamma^\mu j]\over \spa ji}
        & \qquad & 
   \eps_-(\hat j) =  - {z\langle i \gamma^\mu i] - \langle i \gamma^\mu j]
           \over \spb ij} + \cdots  \ .
\eeqa{epsforij}
For the choice $(i,j) = (-,+)$, from the $z$-dependence of the polarization
vectors and the internal vertices and propagators, we see that 
the amplitude cannot be larger than ${\cal O}(1/z)$.  For the choice
$(i,j) = (-,-)$, there is apparently a term of order $z^1$.  However,
it is not possible to build this term, since
\beq
   \eps_-(\hat i) \cdot \eps_-(\hat j) = 0 \quad q \cdot \eps_-(\hat i ) = 
      q \cdot \eps_i(\hat j) = 0 \ .
\eeq{minusminus}
The first nonzero term is one in which both $\eps_-(\hat i)$ and 
$\eps_-(\hat j)$ are dotted with ${\cal O}(1)$  vectors instead of $zq^\mu$,
and this term is down by two powers of $z$, giving in all an amplitude of
of ${\cal O}(1/z)$.  An analogous argument holds for the $(+,+)$ case.
The amplitude in the $(+,-)$ case is irremediably ${\cal O}(z^3)$.

A parallel argument can be carried out for the case in which the shifted
lines are fermions.  In the fermion case, the vertex involves no momenta
and is of order $z^0$, but the propagator is also ${\cal O}(z^0)$.  The
leading term in the propagator is proportional to $\qslash$, which 
annihilates both $i\rangle$ and $j]$.  Then, again, only the $(+,-)$ case
can be nonzero as $z\to \infty$.  There is one exception to this rule:
If we shift on momenta $i$ and $j$ at opposite ends of the {\it same}
fermion line, the diagrams with only a single vertex and no internal 
propagators on this line can give a term of ${\cal O}(z^0)$.

The paper \cite{AHK}, taking a somewhat more sophisticated approach, 
generalizes this conclusion to other theories, including
gravity.

\subsection{Proof of the MHV formula}

As a first application of the BCFW recursion formula, I will give a proof
of the MHV formula \leqn{genMHVg} for $n$-gluon amplitudes, using an argument
originally given by Risager~\cite{Risager}.  The other MHV formulae in 
Section 4 can be proved by the same method.

The proof will proceed by induction.  We have already verified this formula
for the $n = 3,4$ MHV gluon amplitudes.  So, I will 
 assume that the MHV formula
is correct for the case $n = N-1$ and use that hypothesis to evaluate the
$n = N$ gluon MHV amplitude.

\begin{figure}
\begin{center}
\includegraphics[height=2.0in]{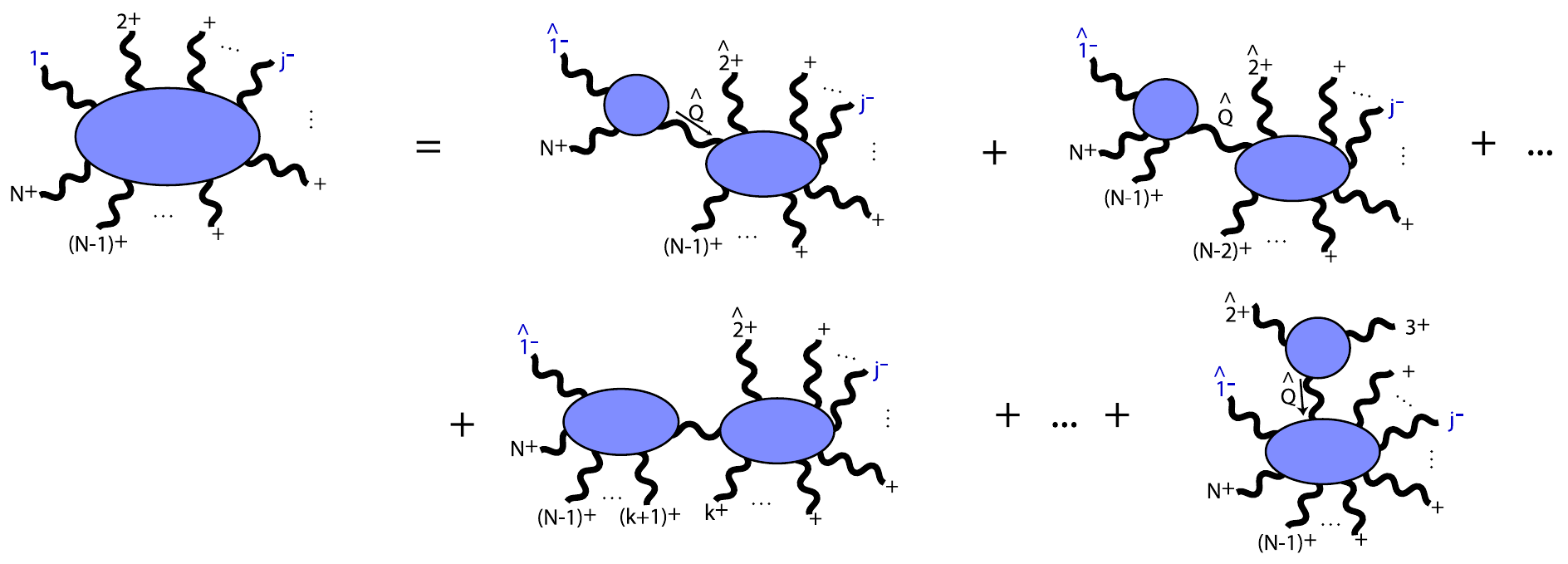}
\caption{BCFW cuts contributing to the evaluation of the $n$-gluon
MHV amplitude.}
\label{fig:MHVBCFW}
\end{center}
\end{figure}

The $N$-gluon amplitudes are cyclically invariant. So, without loss of 
generality, choose the gluon 1 to be one of the two gluons with negative
helicity.  Define the $z$-dependent amplitude in the BCFW procedure by 
the shift 
\beq
  \hat 1] = 1] + z\, 2] \qquad   \hat 2\rangle = 2\rangle - z \, 1\rangle \ .
\eeq{onetwoshift}
The BCFW recursion formula then evaluates the $N$-point MHV amplitude in 
terms of factorized diagrams with $\hat 1$ on one side of the factorization
and $\hat 2$ on the other, as shown in Fig.~\ref{fig:MHVBCFW}.  In each 
diagram, we must assign a helicity to the intermediate gluon.  There are
two choices: $+$ outgoing from one side and $-$ outgoing from the other
side, or vice versa.  In most cases, both choices give zero, since one 
side or the other will be an
 $n$-gluon amplitude with $n > 3$ and one or zero $-$ helicity, which 
vanishes by the arguments given in Section 4.  The two diagrams that
are potentially nonzero are those in which one side or the other is an
amplitude with $n=3$.  These are the first and last diagrams in the 
series shown in Fig.~\ref{fig:MHVBCFW}.

\begin{figure}
\begin{center}
\includegraphics[height=1.2in]{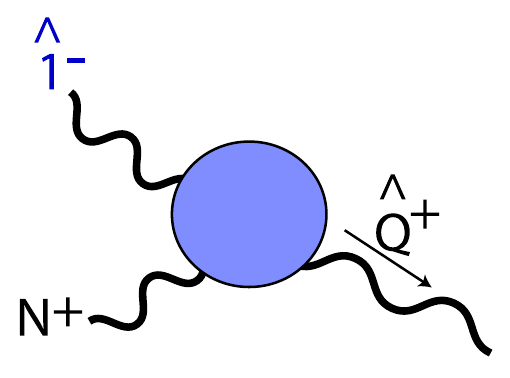}
\caption{The three-point amplitude from the first diagram in 
    Fig.~\ref{fig:MHVBCFW}.}
\label{fig:badthreepoint}
\end{center}
\end{figure}

The first diagram on the right-hand side in the figure involves the 
3-point vertex shown in Fig.~\ref{fig:badthreepoint}.  The denominator of
the exposed propagator is
\beq
   \hat Q^2 =  \spa 1 N \spb N { (1 + z 2)}
\eeq{denominproof}
However, the 3-point vertex has the form
\beq
       -ig   {{ \spb N {\hat Q}}^4 \over \spb {\hat Q}{ \hat 1} \spb {\hat 1}N
       \spb N {\hat Q}}   =  + ig   \spb N {(1+z2)}  \ .
\eeq{vertexinproof}
Then vertex cancels the pole in $z$ from the propagator, so
the diagram actually has no pole.

The only nonzero contribution to the sum then comes from the last diagram.
Its value is
\beq
    i g^{N-3} { {\spa 1 j}^4 (-1) \over \spa 1{\hat Q} \spa {\hat Q} 4 
\spa 45 \cdots \spa N1 } {i\over \spa 23 \spb 32 } (-ig) { {\spb 23}^4
     \over \spb {\hat Q} 2 \spb 23 \spb 3{\hat Q}} 
\eeq{nonzeroinproof}
with
\beq
     \hat Q =  -2\rangle [ 2 - 3 \rangle [3 + z 1\rangle [ 2 \ .
\eeq{Qvalinproof}
From \leqn{Qvalinproof}, we derive
\beqa
    \spa 1 {\hat Q} \spb {\hat Q} 3 &=&  - \spa 12 \spb 23 \CR
    \spa 4 {\hat Q} \spb {\hat Q} 2 &=&  - \spa 43 \spb 32   \ .
\eeqa{extraids}
The factors of $\spb 23$ all cancel, and we end up with the result
\beq
      ig^{N-2} {   {\spa 1j}^4\over \spa 12 \spa 23 \spa 34 \spa 45 \cdots
             \spa N1 } 
\eeq{finalinproof}
which is exactly the Parke-Taylor amplitude for the case of $n=N$ legs.
By induction, this formula applies for all $n$.

The MHV formula for $n$-gluon amplitudes with 2 $+$ helicities, and the 
MHV formulae for amplitudes with 2 fermions and $(n-2)$ gluons can be 
proved by following the same strategy.

\section{$pp \to W^+$ + partons}

As an illustration of this technology, I will now derive the complete 
set of formulae needed to compute the cross section for vector boson
production with up to 2 partons at a hadron collider.  The methods I have
discussed make 
the computation of the cross section for a vector boson plus one parton  
truly trivial.  Only some particular 2 parton amplitudes will require a 
little work.

For definiteness, I will write formulae for $W^+$ production from $u$ and
$d$ quarks.  The contributions from other light flavors, and the corresponding
formulae for $W^-$ and $Z^0$ production, can be derived from these by small
modifications of the prefactors.

\subsection{$u \bar d \to W^+$} 

\begin{figure}
\begin{center}
\includegraphics[height=1.2in]{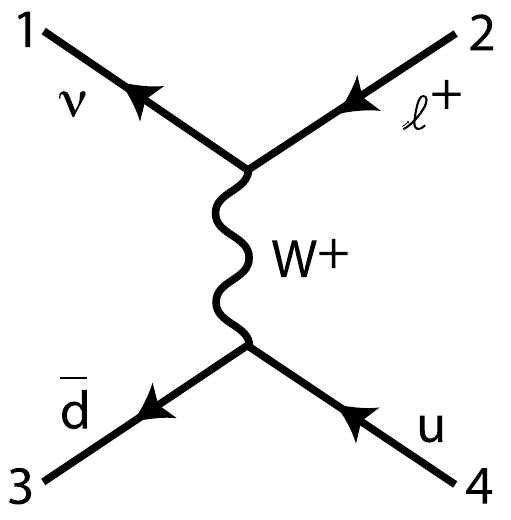}
\caption{Feynman diagrams for $u\bar d \to \nu \ell^+$.}
\label{fig:basicforW}
\end{center}
\end{figure}

The coupling of the $W^+$ to quarks and leptons is 
\beq
\delta \L =  {g_w\over \sqrt{2}} W^-_\mu (\bar d \gamma^\mu P_L u + 
  \bar \ell \gamma^\mu P_L \nu) + h.c. \ ,
\eeq{basicWL}
where $P_L = (1-\gamma^5)/2$ is the left-handed projector, and $g_w$ is the  
weak interaction coupling, satisfying
\beq
     \alpha_w = {g_w^2\over 4\pi} = {1\over 29.6} \ .
\eeq{alphawval}
The $W^-$ field appears because this is the field that creates the $W^+$.
This interaction leads to an amplitude given by the Feynman diagram 
shown in Fig.~\ref{fig:basicforW}
\beqa
   i\M(u\bar d \to \nu \ell^+) &=& i {g_w^2/2
          \over s_{12} - m_W^2 + i m_W \Gamma_W}
      \bar u(1) \bar \sigma^\mu u(2) \, \bar u(3) \bar\sigma_\mu u(4) \CR
&=& i g_w^2 {1 \over s_{12} - m_W^2 + i m_W \Gamma_W} \spa 31 \spb 24 \ .
\eeqa{twototwoforW}
We can treat this formula in one of two ways.  One way is to integrate
over the full phase space of the  
final-state leptons.  This will give a broad mass
distribution for the leptons, on top of which the $W^+$ will appear as a 
resonance.  

Alternatively, since
 the $W^+$ is a narrow resonance, it is typically a good approximation
to compute amplitudes for a real $W^+$ boson 
on its mass shell.  In this case, you might think it 
is easiest to sum over the $W^+$ polarization vectors.
  However, we will obtain simpler formulae if
we retain the final-state lepton spinors.  As a bonus, retaining the spinors
will preserve the information on the $W^+$ polarization.  This is valuable,
first, because the $W^+$ polarization gives useful information on the 
dynamics of the $W^+$ production and, second, because the experimental
acceptance for a $W^+$ depends strongly on its polarization.

The matrix element for the $W^+$ coupling to leptons is proportional to 
the current $\bar u(\nu) \bar\sigma^\mu v(\ell^+)$.  This matrix element
must be squared and integrated over the direction of the leptons in the 
$W^+$ rest frame. That  integral is
\beq
    I^{\mu\nu} = 
 \int {d\Omega\over 4\pi} \  \langle 1 {\gamma^\mu} 2]\langle 2 \gamma^\nu 1]
         \ .
\eeq{WOmega} 
To evaluate this integral, note that, if $q^\mu$ is the $W^+$ momentum, 
$q = 1+2$, then $q_\mu I^{\mu\nu} = q_\nu I^{\mu\nu} = 0$.  Further, 
\beq
    I^{\mu}{}_\mu = 
 \int {d\Omega\over 4\pi} \   2 \spa 12 \spb 12 = - 2 q^2 = - 2 m_W^2 \ .
\eeq{Itrace}
From these requirements,
\beq
   I^{\mu\nu} =  - {2\over 3}m_W^2 (g^{\mu\nu} - {q^\mu q^\nu\over m_W^2}) \ .
\eeq{Ifinal}
The quantity in parentheses in this equation is the usual sum over 
on-shell $W^+$ polarization vectors.  Thus we can represent this sum as 
\beq
       \sum_i \eps_i^\mu(q) \eps_i^{*\nu}(q)  = {3\over 2m_W^2}
\int {d\Omega\over 4\pi} \  \langle  1 {\gamma^\mu} 2]
           \langle 2 \gamma^\nu 1]\ .
\eeq{epssumrep}

This gives a simple procedure for computing cross sections with a final 
on-shell $W^+$~\cite{KS}:   (1) Write the Feynman diagrams with an 
internal $W^+$ propagator and the final state $\nu(1)\ell^+(2)$.
(2) Remove the factor 
$(g_w/\sqrt{2})/(s_{12} - m_W^2 + i m_W\Gamma_W)$ and put the
$W^+$ momentum (1+2) on shell. \ (3) Evaluate the amplitude
with spinor products.  (4)  Square and integrate over
phase space including
the on-shell $W^+$, and add an integral over the lepton direction in the 
$W^+$ rest frame:
\beq
  \int d\Pi_{n+W} \equiv
       \int d \Pi_{n+1}  {3\over 2 m_W^2} \int {d\Omega\over 4\pi}\ .
\eeq{newPiint}
The cross section for $W^+$ production is typically
 already a multi-dimensional Monte Carlo integral, so the computational price
of adding two additional, well-behaved integrals is small.   

To illustrate this formalism, I will work out the cross section for 
$u\bar d \to W^+$.  I use the matrix element \leqn{twototwoforW} in the
form
\beq 
 i\M(W^+ d\bar u) =  {g_w\over\sqrt{2}} \langle 1 \gamma^\mu 2]\langle
   3 \gamma^\mu 4] =  \sqrt{2} g_w  \spa 31 \spb 24 \ .
\eeq{newmatrixudW}
The phase space integral $\int d\Pi_{0+W}$ contains 1-body phase
space
\beq
        \int d\Pi_1 =   2\pi\delta(s - m_W^2) \ .
\eeq{onebodyps}
Then, averaging over initial colors and spins, we find
\beq
   \sigma(u\bar d \to W^+) = {1\over 3\cdot 4} \cdot
 {8\pi \alpha_w\over 2s}\cdot 2\pi\delta(s - m_W^2) \cdot
 {3\over 2 m_W^2} \int {d\Omega\over
    4\pi} \   s_{24}^2 \ .
\eeq{udWcsone}
The integral over $d\Omega$
implements the familiar $(1-\cos\theta)^2$ angular 
distribution of the decay lepton with respect to the $u$ quark direction,
signalling a $W^+$ with left-handed polarization.  Evaluating the 
integral, we find
\beq
   \sigma(u\bar d \to W^+) = {\pi^2 \alpha_w\over 3}\delta(s - m_w^2) \ ,
\eeq{DYsigma}
the familiar expression for the Drell-Yan cross section.

\subsection{MHV amplitudes}

To use this formalism for computing $W^+$ production cross sections, we need
to be able to compute the amplitudes.  We might hope that many of these
amplitudes belong to an MHV series and are therefore trivial to evaluate.
In fact, it is so.

\begin{figure}
\begin{center}
\includegraphics[height=1.7in]{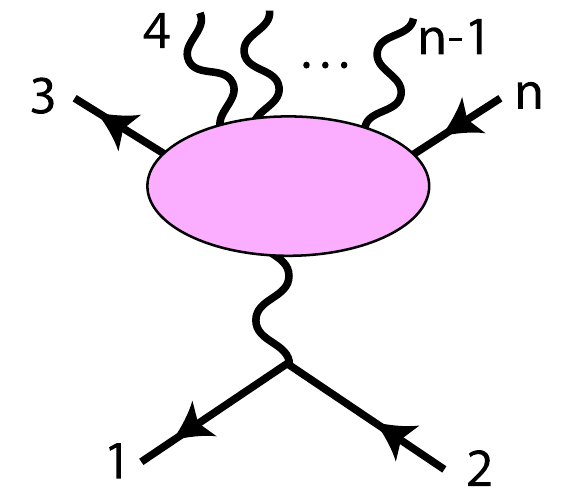}
\caption{Notation for the amplitudes for $e^+_Re^-_L \to q_L + ng + \bar q_R$.}
\label{fig:eeqgggq}
\end{center}
\end{figure}

To explain this, I will first compute some amplitudes for the related process
$\ee\to q \bar q + n g$.  We have seen the first of these, for zero gluons,
 already in 
\leqn{allangle} and \leqn{allsq}.  This amplitude seems to belong to both
an MHV and an anti-MHV series.  For convenience, I will renumber the external
legs to correspond better to the formalism that we used in Section~4.  On 
the electron side, I will take the outgoing left-handed line (the incoming 
$e^+_R$) to be line 1; on the quark side, I will take the outgoing left-handed
line (the outgoing $q_L$) to be line 3.  I will notate a general amplitude
for $e^+_R e^-_L \to q_L + n g + \bar q_R$ in the same way, as shown in 
Fig.~\ref{fig:eeqgggq}.  Then the $n = 0$ case gives
\beq
  i\M =  -2ie^2{ { \spa 13 }^2 \over \spa 12 \spa 34} = -2ie^2 { { \spb 24}^2
      \over \spb 12 \spb 34 } \ .
\eeq{zerogluoncase}
The electric charges gives an additional factor $(-Q_f)$;
I omit this here and in the following.

It is not difficult to work out the next case, $n=5$ or 1 gluon, using the 
methods described in this review.  In the case of a $+$ helicity gluon, it 
is easiest to take the reference vector for the gluon (line 4) to be 3; for a 
$-$ helicity gluon, take the reference vector to be 4.  Then one finds
\beqa
  i\M(e_-(1) \bar e_+(2) q_-(3)  g_+(4) \bar q_+(5) ) 
   & =&  -2ie^2 g_s { { \spa 13 }^2 \over \spa 12 \spa 34 \spa 45} \CR
  i\M(e_-(1) \bar e_+(2) q_-(3)  g_-(4) \bar q_+(5) ) 
   & =&  2ie^2 g_s { { \spb 24 }^2 \over \spb 12 \spb 34 \spb 45}\ ,
\eeqa{onegluoncase}
times the factor $(-Q_f)$ and the color matrix $T^a$.
The sum of the squares of these expressions gives
\beq
\sum |\M|^2 = 4e^4 g_s^2 \ {s_{13}^2 + s_{24}^2 \over s_{12} s_{34} s_{45} }\ .
\eeq{ssquareeeqqg}
This reaction is usually described using the kinematic variables
\beq
  x_q = {2 k_3\cdot q\over q^2} \ , \quad  x_g = {2 k_4\cdot q \over q^2}\ ,
          \quad
      x_{\bar q} = {2 k_5 \cdot q \over q^2} \ ,
\eeq{xdefin}
where $q = 3+4 +5 = -(1+2)$.  Each $x_i$ is the ratio of that particle's
momentum in the center of mass frame to its maximum value $q^0/2$.  Also,
\beq
   s_{34} = q^2 (1-x_{\bar q})\ , \quad s_{45} = q^2 (1-x_q) \ , \quad
  s_{35} = q^2 (1-x_{g})\ .
\eeq{svalues}
The standard expression for the cross section for gluon production in 
$\ee$ is
\beq
   {d \sigma\over d x_q dx_{\bar q} } =  { 4\pi\alpha^2 \over 3 s}\cdot 
   3Q_f^2 \cdot 
 {2\alpha_s\over 3 \pi}\cdot { x_q^2 + x_{\bar q}^2 \over (1-x_q)(1
        - x_{\bar q})} \ .
\eeq{standardg}
The last factor is almost exactly of the form of \leqn{ssquareeeqqg} after
substituting \leqn{xdefin}, \leqn{svalues}.  By reintroducing $Q_f$, summing
over final colors, averaging over helicities, and averaging over the relative
orientation of the final state plane with respect to the beam axis, it is 
not difficult to complete the derivation of \leqn{standardg} from 
\leqn{ssquareeeqqg}.

It is tempting to guess that the formulae \leqn{onegluoncase} generalize
to the color-ordered amplitudes for 
all $+$ helicity and all $-$ helicity gluons according to 
\beqa
 & &  i\M(e_-(1) \bar e_+(2) q_-(3)  g_+(4)\cdots g_+(n-1) \bar q_+(n) ) \CR
  &  & \hskip 2.0in =  - 2ie^2 g_s^{n-4} { { \spa 13 }^2 \over \spa 12 
 \spa 34 \spa 45 \cdots \spa {(n-1)}n } \CR
& &   i\M(e_-(1) \bar e_+(2) q_-(3)   g_-(4)\cdots g_-(n-1) \bar q_+(n) )\CR 
   & & \hskip 1.6in = (-1)^{n+1} 2ie^2 g_s^{n-4}
   { { \spb 2n }^2 \over \spb 12 \spb 34 \spb 45 \cdots 
\spb {(n-1)}n }\ .
\eeqa{ngluoncase}
This is in fact correct.  These new MHV formulae can be proved using the 
method of Section 6.3.

Finally, we need to make the connection between the QED amplitudes given 
here and the amplitudes that we need for $W^+$ production.  The
only change needed is to substitute the photon propagator with a $W^+$ 
propagator by  multiplying by the factor 
\beq
   {g_w^2\over 2 e^2}{   s_{12} \over s_{12} - m_W^2 + i m_W \Gamma_W} \ , 
\eeq{subsprop}
and then to take the $W^+$ on-shell according to the prescription given
below \leqn{epssumrep}.  I will write the on-shell amplitudes in the form
\beqa
  i\M(W^+(12);3\cdots n) &=&  
     \sqrt{2} g_w  g_s^{n-4} T^{a_{1}} \cdots T^{a_{n}} \cdot
         i\bfM(12;34\cdots (n+4)) \CR  & & \hskip 2.0in
      +   \mbox{other\ color\ structures} \ .
\eeqa{Wform}
Then for the cases discussed above,
\beq
 \bfM(12;3\cdots n) = \cases{\phantom{(-1)^n}
 {\spb 12 { \spa 13 }^2 / \spa 12 
 \spa 34 \spa 45 \cdots \spa {(n-1)}n }  & MHV \cr 
     (-1)^n {\spa 12 { \spb 2n }^2 / \spb 12 
 \spb 34 \spb 45 \cdots \spb {(n-1)}n } &  anti-MHV \cr} \ .
\eeq{MHVcaseofW}

\subsection{$W^+$ + 1 parton}

The amplitude for $u\bar d \to W^+ + g_+$ contains only one color 
factor.  The corresponding color-ordered amplitude, following from 
\leqn{onegluoncase} or \leqn{MHVcaseofW}, is
\beq
  \bfM =   {\spb 12 {\spa 13}^2 \over \spa 34 \spa 45} \ .
\eeq{exampleW}
Using this result, it is straightforward to compute the lowest-order
cross section  for production of $W^+$ + 1 parton.  There are three 
processes that must be considered:
\beq
   u \bar d \to W^+ g  \ , \quad    u  g \to W^+ d \ , \quad
g \bar d \to W^+ \bar u
        \ .
\eeq{threeprocs}
For all three processes, the 
amplitude is an appropriate crossing of that given in \leqn{exampleW}.

For $u\bar d \to W^+ g$, after
averaging over initial colors and helicities 
and summing over final colors and helicities, we find
\beqa
\sigma(  u \bar d \to W^+ g) &=& {32\over 9} {\pi^2 \alpha_w\alpha_s\over s}
     \int d\Pi_{1+W} \  \CR  & & \hskip 0.05in \biggl\{  
 |\bfM(12;d(3) g_+(4) \bar u(5)|^2 +
 |\bfM(12;d(3) g_-(4) \bar u(5)|^2 \biggr\} \ .
\eeqa{onegcross}
I have used the color sum
\beq
          \tr [ T^a T^a ] = 8  \ .
\eeq{traceforthis}
If  $\theta_*$ is the polar angle in the $W^+ g$ center of mass
system,
\beq
   \int d\Pi_{1+W} =  {3\over 16\pi m_W^2} 
\int {d\cos\theta_*\over 2} \int {d\Omega\over
       4\pi}  \bigl( 1 - {m_W^2\over s}\bigr) 
\eeq{twoWphasespace}
with $s = s_{5}$, and 
\beq
 |\bfM(12;d(3) g_+(4) \bar u_+(5)|^2 +
 |\bfM(12;d_-(3) g_-(4) \bar u_+(5)|^2 =  { s_{13}^2 + s_{25}^2 \over 
         s_{34} s_{45}} \ .
\eeq{twoWmatrix}
Then
\beq
\sigma(u\bar d \to W^+g) = {2\pi \alpha_w\alpha_s\over 3s} 
\int {d\cos\theta_*\over 2} \int {d\Omega\over
       4\pi}  (1 - {m_W^2\over s}) \biggl( {s_{1d}^2 + s_{2u}^2\over s_{dg}
       s_{gu}}\biggr)\ .  
\eeq{ubardtoWg}

Similarly,
\beq
\sigma(u g \to W^+ d ) = \sigma(  g\bar d \to W^+ \bar u )
=  {\pi \alpha_w\alpha_s\over 4  s}
\int {d\cos\theta_*\over 2} \int {d\Omega\over
       4\pi}  (1 - {m_W^2\over s}) \biggl( {s_{1d}^2 + s_{2u}^2\over s_{dg}
       s_{gu}}\biggr)\ .  
\eeq{ugtoWd}
where the $s_{ij}$ invariants are evaluated with appropriately crossed
momenta.  In the two cases, respectively, $s_{gu} = s$ and $s_{dg} = s$.

\subsection{Cross section formulae for $W^+$ + 2 partons}

For the case of $W$ + 2 partons, many reactions must be included, and 
each of these has a nontrivial color structure.  The reactions are all
of the form (2 partons) $\to$ (2 partons) $+ W^+$, and so they are in 
1-to-1 correspondence with the 2-to-2 parton processes that we considered
in Section 5.  Further, since the $W^+$ is a color singlet, the color 
structure and color sums are also just those that we met in Section 5,
and we can borrow the color algebra from that discussion.
This will allow us to write the cross sections in terms of the matrix 
elements $\bfM$ normalized as in \leqn{Wform}.

We can begin with 4-fermion reactions.  For the simplest case of a 
non-identical spectator quark, the cross section is given in terms of
the amplitude $\bfM(12;34;56)$ containing three fermion lines: the leptons
(12), the left-handed quark emitting the $W^+$ (34), and a second left-handed
quark (56).  This cross section is
\beq
\sigma(us\to W^+ ds) = {1\over 9} {\alpha_w\alpha_s^2\over s}
    \cdot 128\pi^3\, \int d\Pi_{2+W} \
\biggl\{ |\bfM(12;34;56)|^2 + |\bfM(12;34;65)|^2 \biggr\} \ ,
\eeq{sigusds}
The sum of matrix elements takes care of the polarization sum for the 
spectator quark.  For spectators identical to the 
final quark,
\beqa
\sigma(ud\to W^+ dd) &=& {1\over 9} {\alpha_w\alpha_s^2\over s}
   \cdot 128\pi^3\, \int d\Pi_{2+W} \CR & & \hskip -0.1in
 \biggl\{ |\bfM(12;34;56)|^2  - {1\over 3}
       \Re[ \bfM(12;34;56)^* \bfM(12;54;36)] \CR
       & & \hskip 1.0in + |\bfM(12;34;65)|^2]\biggr\}\ .
\eeqa{siguudu}
I have adjusted the normalization so that the integral can be taken
over all of phase space.  The cross section $\sigma(uu\to W^+ d u)$ 
involves a similar sum over matrix elements.
Cross sections involving antiquarks have the same form, except that the 
amplitudes $\bfM(12;34;56)$ should be crossed appropriately.

The cross section for the reaction $u\bar d \to W^+ + 2g$ is given, analogously
to \leqn{uuggfinal}, by 
\beqa
\sigma(u\bar d\to W^+ gg)& = &{8\over 27} {\alpha_w\alpha_s^2\over s}
    \cdot 128\pi^3\,  \int d\Pi_{2+W} \sum_{ij} \CR
& & \hskip -0.4in \biggl\{ |\bfM(12;34_i5_j6)|^2 
- {1\over 8} \Re[ \bfM(12;34_i5_j6)^* \bfM(12;35_j4_i6)]\biggr\}\ ,
\eeqa{uuggWfinal}
where $i,j = +,-$ are the helicity states of the final gluons.  The
cross section is again normalized to be integrated over all of phase
space.  The
cross sections 
with initial-state gluons are given by 
\beqa
\sigma(ug\to W^+ ug) &=& {1\over 9} {\alpha_w\alpha_s^2\over s}
     \cdot 128\pi^3\, \int d\Pi_{2+W} \sum_{ij}\CR
& & \hskip -1.0in \biggl\{ |\bfM(12;34_i5_j6)|^2 +  |\bfM(12;35_j4_i6)|^2 
- {1\over 4} \Re[ \bfM(12;34_i5_j6)^* \bfM(12;35_j4_i6)]\biggr\} \CR
\sigma(gg\to W^+ d\bar u) &=& {1\over 24} {\alpha_w\alpha_s^2\over s}
     \cdot 128\pi^3\, \int d\Pi_{2+W} \sum_{ij}\CR
& & \hskip -1.3in \biggl\{ |\bfM(12;34_i5_j6)|^2  +  |\bfM(12;35_j4_i6)|^2 
- {1\over 4} \Re[ \bfM(12;34_i5_j6)^* \bfM(12;35_j4_i6)]\biggr\}\ ,
\eeqa{otheruuggWfinal}
where the matrix elements $\bfM(12;3456)$ are crossed appropriately.

There is no analogue of the $gg\to gg$ process, since at tree level the $W^+$
does not couple directly to gluon lines.  So now we have all of the 
cross section formulae, and it only remains to compute the amplitudes.

\subsection{Amplitudes for $W^+$ + 2 partons}

The cross section formulae in the previous section involve two sets of 
amplitudes, those for $Wqgg\bar q$ and those for $Wq\bar qq\bar q$. These must
be evaluated for all cases of the final-state helicity.  However, each
quark line must have a left-handed fermion at one end and a right-handed
fermion on the other end,
so the cases that must be considered are only those of different 
gluon helicities.  Of these, two are MHV or anti-MHV, so only two amplitudes
with gluons, plus the four-fermion case, need to be computed.  All of those
computations are easily done by the BCFW method.

The two amplitudes that are trivially known are:
\beqa
  \bfM(12;d_-(3)g_+(4)g_+(5) \bar u_+(6))&=&  { \spb 12{\spa 13}^2
      \over \spa 34 \spa 45 \spa 56} \CR
  \bfM(12;d_-(3)g_-(4)g_-(5) \bar u_+(6))&=& { \spa 12  {\spb 26}^2
      \over \spb 34 \spb 45 \spb 56}  \ .
\eeqa{knownMs}

\begin{figure}
\begin{center}
\includegraphics[height=1.5in]{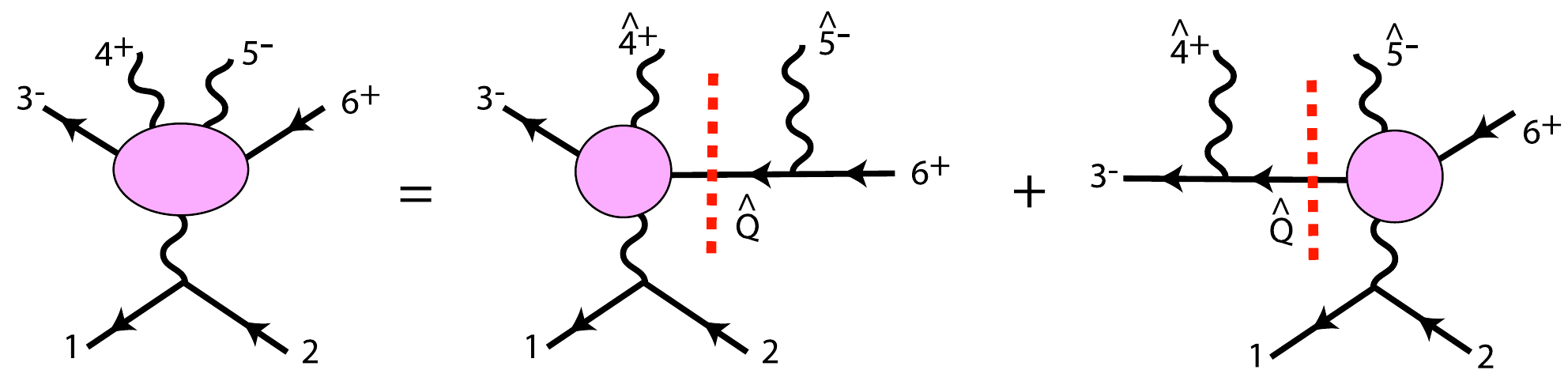}
\caption{Evaluation of the amplitude $i\bfM(12;d_-(3)g_+(4)g_-(5)\bar u_+(6))$
 by BCFW recursion.}
\label{fig:sixID}
\end{center}
\end{figure}

I will now take up the case of the amplitude 
$i\bfM(12;d_-(3)g_+(4)g_-(5)\bar u_+(6))$.  To evaluate this amplitude, 
carry out a 
BCFW shift on the legs 5 (with square brackets) and 4 (with angle 
brackets, according to 
\beq
        \hat 5] = 5] + z\ 4] \ , \qquad   \hat 4\rangle = 4\rangle - z\ 
                5 \rangle \ .
\eeq{shiftforsix}
Then the BCFW recursion formula gives the identity shown in 
Fig.~\ref{fig:sixID}.  The vertical dashed lines show the two BCFW cuts.

I will evaluate the first of these cuts explicitly.  The amplitude to the 
left is an MHV amplitude for $Wug\bar d$.  The amplitude to the right
is a nonzero MHV 3-point vertex.  Together, these give
\beq
  {\spb 12 {\spa 13}^2\over \spa 3 {\hat 4} 
       \spa {\hat 4}{(-\hat Q)}} \cdot(-i) \cdot {i \over s_{56} } \cdot 
 i { {\spa {\hat Q}{\hat 5}}^3 \spa 6{\hat 5} \over \spa {\hat Q}{\hat 5}
       \spa {\hat 5} 6 \spa 6 {\hat Q}} \ .
\eeq{BCFWfirst}
The factor $(-i)$ is associated with the fermion cut; see below 
\leqn{fermionprop}.  Using $\hat 5\rangle = 5 \rangle$ and 
$-Q\rangle = i Q\rangle$, this simplifies to 
\beq
   -  {\spb 12 {\spa 13}^2 {\spa 5{ \hat Q}}^2 \over  \spa 3 {\hat 4} s_{56}
       \spa {\hat 4}{\hat Q}  \spa 6 {\hat Q} } \ .
\eeq{simplerBCFWfirst}

The momentum $\hat Q$ takes the value
\beq
   \hat Q = \hat Q\rangle [ \hat Q = 
      -  5\rangle[5 - 6 \rangle [ 6 - z\ 5\rangle [ 4 \ .
\eeq{Qvalue}
This should be evaluated at the value $z = z_*$ for which $\hat Q^2 = 0$.
That is,
\beq
    z_* = - {s_{56}\over \langle 5 (5+6) 4] } = - {\spb 65\over \spb 64} \ .
\eeq{zstarval}
To evaluate the angle brackets of $\hat Q$, multiply \leqn{simplerBCFWfirst}
by ${\spb {\hat Q} 4}^2/{\spb {\hat Q} 4}^2$ and use \leqn{Qvalue} to 
evaluate $\hat Q\rangle [ \hat Q$.  This gives the 
identities
\beqa
   \spa 5 {\hat Q} \spb {\hat Q}4 &=&  - \spa 5 6 \spb 64 \CR
   \spa 6 {\hat Q} \spb {\hat Q}4 &=&  - \spa 65 \spb 54 \CR
   \spa 4 {\hat Q} \spb {\hat Q}4 &=&  - (s_{45} + s_{46} + s_{56}) = - s_{456}
\eeqa{Qhatids}
Also, using \leqn{zstarval},
\beq
    \spa 3 {\hat 4} =  \langle 3 (4 +5) 6]/ \spb 46 \ .
\eeq{lastid}
With all of these simplifications, the expression for the cut becomes
\beq
  { \spb 12 {\spa 13}^2 {\spb 46}^3\over \spb 45 \spb 56
           s_{456} \langle 3 (4+5) 6] } \ .
\eeq{lastsimple}   
   
After evaluating the second cut in the same way, we arrive at the following
expression for the amplitude:
\beq
 \bfM(12;d_-(3)g_+(4)g_-(5) \bar u_+(6)) = 
 \biggl\{ 
  { \spb 12 {\spa 13}^2 {\spb 46}^3\over \spb 45 \spb 56
           s_{456} \langle 3 (4+5) 6] } +
  { \spa 12 {\spb 26}^2 {\spa 35}^3\over \spa 34 \spa 45
           s_{345} \langle 3 (4+5) 6] } \biggr\} \ .
\eeq{finalsecond}
The final result is properly symmetric under complex conjugation and 
interchange of $3\leftrightarrow 6$, $4\leftrightarrow 5$.  Notice that
the separate terms of the amplitude are singular not only at physical 
singularities where
 denominators vanishes, such as $s_{45} = 0$, but also 
on the plane where $\langle 3 (4+5) 6] = 0$.  Fortunately, this latter
singularity can be shown to cancel between the two terms of the expression.

The two remaining amplitudes can be evaluated by the same technique. For
the remaining $Wugg\bar d$ amplitude, shifting on 4 and 5 gives
\beq
 \bfM(12;d_-(3)g_-(4)g_+(5) \bar u_+(6)) = 
 \biggl\{ 
  { \spa 12 {\spb 46}{ [ 2 (5+6)4\rangle}^2 \over \spa 45 \spa 56
           s_{456} [ 3 (4+5) 6\rangle } +
  { \spb 12 \spb 35 {\langle 1 (3+4) 5 ]}^2 \over \spb 34 \spb 45
           s_{345} [ 3 (4+5) 6\rangle  } \biggr\} \ .
\eeq{finalmp}
For the 4-fermion amplitude, shifting on 3 and 6 gives
\beq
  \bfM(12;d_-(3)\bar u_+(4);s_-(5) \bar s_+(6))  = - \biggl\{
  { \spa 12 {\spb 24}^2 {\spa 35}^2 \over \spa 56
           s_{356} [ 4 (5+6) 3\rangle } -
  { \spb 12 {\spa 13}^2 {\spb 46}^2\over \spb 56
           s_{123} [ 4 (5+6) 3\rangle  } \biggr\} \ .
\eeq{finalfermion}
In this case, one should resist the temptation to shift on the lines 
5 and 6.  That gives a simpler expression, which, however, is not actually
correct,
because the limit $z \to\infty$ in the BCFW construction 
does not vanish.   Alternatively, the 4-fermion 
amplitude can be computed very simply directly from the Feynman diagrams
using only the technology 
of Section 2.  The result is
\beq
  \bfM(12;d_-(3)\bar u_+(4);s_-(5) \bar s_+(6))  = - \biggl\{
  { \spa 35 [ 6 (3+5) 1 \rangle \spb 24 \over s_{56} s_{356}} 
+  { \spa 31 [ 2 (1+3) 5 \rangle \spb 64 \over s_{56} s_{123}} 
 \biggr\} \ .
\eeq{finalfermiontwo}
which can be shown to be equal to \leqn{finalfermion}.

In principle, one could go further to amplitudes with $W^+$ and any 
number of quarks and gluons.  In practice, it is possible to avoid
even writing these expression explicitly.  It is quite straightforward to
implement the BCFW recursion in a recursive computer algorithm~\cite{Weinz}.
  Then
these and higher expressions for amplitudes would be generated automatically
starting from the MHV and anti-MHV amplitudes.   Alternatively, these
higher point amplitudes are computed in~\cite{HennDix,Bourj}.

\section{Conclusions}

In these lectures, I have given an introduction to a set of tools for 
computing multi-parton tree level amplitudes in QCD.   Already at the
level of this review, we have seen that calculations that would be 
tremendously difficult by textbook methods become accessible, or even 
easy, using the simplifications of spinor products, color ordering, and
BCFW recursion. For loop diagrams, even more powerful methods are available,
as you might glean from the references.  I hope that this review will put 
you on the road to a better understanding of QCD that will be useful to you
in the era of LHC physics.

\appendix

\section{Direct computation of spinor products}

In numerical computations with spinor products, it is best to evaluate
the spinor products directly from 4-vectors, without first computing 
vector products. The advantage is not only in speed of execution.
 In the case where two lightlike vectors are almost
collinear, separated by a small angle $\theta$, the vector product 
vanishes as $\theta^2$ while the spinor 
product vanishes only as $\theta^1$.  In such a case, working directly with
the spinor product avoids round-off error~\cite{KS}.

To evaluate the spinor products of lightlike vectors A and B, first let
\beq
   \eta_A = \cases{ 1 &  $A^0 > 0$ \cr    -i  &  $A^0 < 0$ \cr }
\eeq{etadef}
Let 
\beq
      A^+ = A^0 + A^1  \qquad    B^+ = B^0 + B^1 \ .
\eeq{plusdefin}

\beq
\spa AB = \eta_A \eta_B { (A^3 - i A^2) B^+ - (B^3 - i B^2) A^+ \over
       \sqrt{ \eta^2_A \eta^2_B A^+ B^+} }
\eeq{spaAB}
and  $\spb AB = - (\spa AB)^*$.

In this definition, the axis $\hat 1$ has a preferred role. In compensation,
  the spinor 
product has a singularity in its phase when one of the vectors $A$, $B$
approaches the $-\hat 1$ direction.  This phase choice cancels out of
the squares of matrix elements, so any other axis may be used.
If the preferred axis is a natural axis of the problem such as
the beam direction, one will frequently encounter 0/0 in numerical evaluations.
If the beam axis is the $\hat 3$ direction, it is useful to choose
 the $\hat 1$ direction
as the preferred axis.  Then, working in double precision, excessively small
denominators appear only for about 1 point in $10^6$; in a Monte Carlo 
integration, one can trap for and ignore these points.

\Acknowledgements

I am grateful to Maria Elena Tejeda-Yeomans and Alejandro Ayala for the 
invitation to present this material in Sonora and for their kind 
hospitality.  I thank
Yu-Ping Kuang, Qing Wang, and Hong-Jian He for the invitation to speak at 
Tsinghua University and for gracious
 hospitality during my stay in Beijing.  I am 
grateful to Zvi Bern, Sally Dawson, and  Lance Dixon, and Daniel Maitre
 for discussions of these
topics.  I thank the many students who have helped me understand this
material and have critiqued my lectures, in particular, 
 Kassahun Betre,
Camile Boucher-Veronneau, Shao-Feng Ge, 
Xiomara Gutierrez, Andrew Larkoski, and My Phuong Le, 
  This 
work was supported by the US Department of Energy under contract 
DE--AC02--76SF00515.

\end{document}